\documentclass[aps,pra,twocolumn,floatfix,showpacs]{revtex4-2}
\usepackage{hyperref}
\usepackage[T1]{fontenc}
\usepackage[utf8]{inputenc}
\usepackage{amsmath}
\usepackage{amssymb}
\usepackage{esint}
\usepackage{color}
\usepackage{graphicx}
\usepackage{epsfig}
\usepackage{pstricks}
\usepackage{physics}
\usepackage{multirow}
\usepackage{float}
\usepackage{diagbox}
\usepackage{array}
\usepackage{hhline}
\usepackage{tabularx}
\usepackage[normalem]{ulem}
\usepackage{amsfonts}
\usepackage{dcolumn}
\usepackage{bm}
\usepackage{overpic}

\newcommand{\e}[1]{\times 10^{#1}}
\newcommand{\bvec}[1]{\boldsymbol{\mathbf{#1}}}
\newcommand{\mat}[1]{\boldsymbol{\mathbf{#1}}}
\newcommand{\uvec}[1]{\hat{\boldsymbol{\mathbf{#1}}}}
\newcommand{\bmat}[1]{\begin{bmatrix}#1\end{bmatrix}}
\newcommand{\HePlus}{{$\text{He}^{+}\,$}}
\newcommand{\I}{\mathrm{i}}
\newcommand{\E}{\mathrm{e}}
\newcommand{\leftgroup}[1]{\left\{\begin{aligned}#1\end{aligned}\right.}
\newcommand{\Eqref}[1]{(\ref{eq:#1})}
\newcommand{\Figref}[1]{Fig.~\ref{fig:#1}}

\graphicspath{{figures/}}

\begin{document}

\preprint{}

\title{Photoelectron -- residual-ion entanglement in streaked shake-up ionization of helium}

\author{Hongyu Shi and Uwe Thumm}

\affiliation{Department of Physics, Kansas State University,
Manhattan, Kansas 66506, USA}

\date{\today}

\begin{abstract}

Streaked photoelectron emission spectra access the correlated dynamics of photoelectrons and residual target electrons with attosecond temporal resolution. We calculated {\em ab initio} single-ionization spectra for photoemission from helium atoms by co-linearly polarized ultrashort XUV and assisting few-femtosecond IR pulses.
Distinguishing direct and shake-up ionization resulting in ground-state and excited ($n=2,3$) \HePlus residual ions, respectively, we examined the effects of the correlated photoemission dynamics on the photoelectron phase-accumulation as a function of the observable photoelectron detection direction and kinetic energy, and XUV -- IR pulse delay. We tracked the dynamical evolution of the residual ion in relative streaked photoemission delays and found dominant contributions for shake-up emission from the residual ion -- photoelectron interaction. 
These are in very good and fair agreement, respectively,  for $n=2$ and $n=3$ shake-up  photoemission along the pulse-polarization directions, with previous experimental and theoretical investigations [M. Ossiander {\em et al}, Nature Phys 13, 280–285 (2017)] and reveal a strong photoemission-direction dependence for shake-up ionization due to the coupling between the photoelectron and evolving residual-ion charge distribution in the IR-laser field. 
\end{abstract}

\maketitle

\section{Introduction} \label{sec:intro}

\begin{figure}[h]
\centering{}\includegraphics[width=1\linewidth]{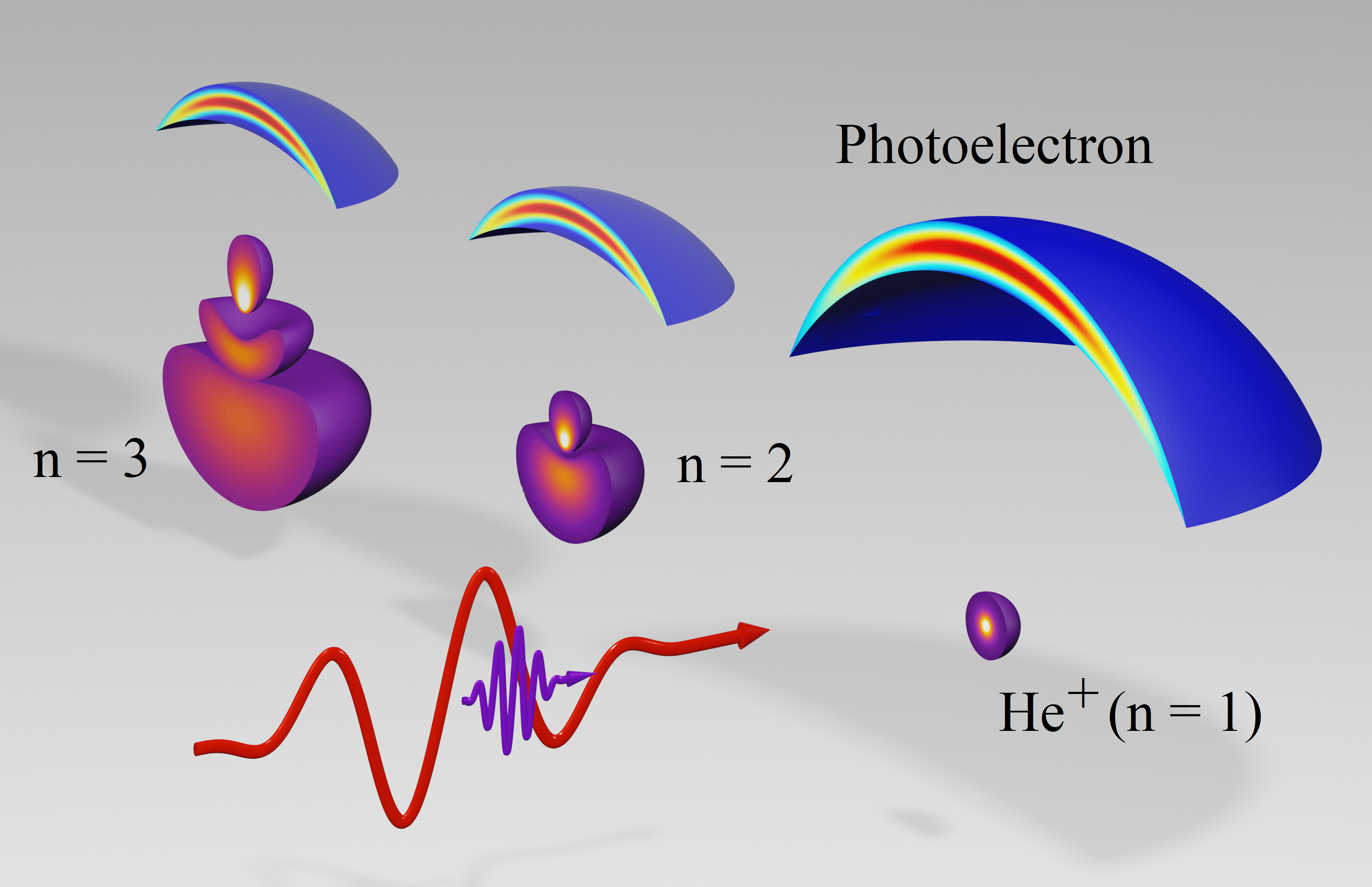}
\caption{Streaked direct (principal quantum number $n=1$) and shake-up ($n=2, 3$) single ionization of helium. Illumination of ground-state helium atoms by ionizing attosecond XUV pulses (purple) and assisting delayed few-femtosecond streaking IR pulses (red) create entangled pairs of PE wave packets and residual \HePlus ($n$) ions. The displayed PE-wave-packet probability densities are calculated {\em ab initio} 4~fs after the XUV-pulse center has reached the atomic nucleus for an XUV-IR pulse delay of -1/4 IR period. The residual \HePlus ($n=2,3$) charge densities for the most polarized Stark states
and pulse amplitudes and wavelengths are sketched and not to scale.}
\label{fig:art3D}
\end{figure}

Attosecond (1 as = 10$^{-18}$~s) streaking is an experimental technique for investigating the time-resolved ultrafast electronic dynamics in atoms~\cite{Schu10,Ossi17}, molecules~\cite{Quan19,Catt22,Kowa16}, and solids~\cite{Cava07,Forg16,Ossi18}. Attosecond streaking experiments provided the first direct measurement of the oscillating electric field in intense light pulses and confirmed the generation of ultrashort sub-infrared (IR) cycle extreme ultraviolet (XUV) pulses with attosecond duration~\cite{Hent01,Goul04}. In a typical streaking experiment, an ultrashort XUV-pump pulse ionizes the target, emitting photoelectrons (PEs) into the field of a femtosecond infrared (IR) probe pulse  (\Figref{art3D}). This allows the recording of PE energy (or momentum) spectra for a set range of time delays $\tau$ between the two pulses. The $\tau$-dependent spectra reveal information about the photoemission dynamics with attosecond resolution in time as "streaking traces", i.e., as an oscillating PE-energy (or momentum) dependence on $\tau$ that retraces the temporal profile of the IR streaking pulse with a characteristic {\em absolute} photoemission time delay relative to the IR-laser carrier electric field. While absolute streaking delays cannot be measured directly for photoemission from gaseous targets~\cite{Ossi18}, dynamical information can be accessed in streaked PE spectra from {\em relative} streaking delays between the centers of energy of streaking traces that are related to photoemission from two energetically distinct electronic levels~\cite{ThummChap15}. 

{\em Relative} streaked photoemission time delays of $21\pm 5$~as were measured between photoemission from the 2$s$ and 2$p$ orbitals of neon by Schultze {\em et al.}~\cite{Schu10} in a milestone experiment that triggered an intense debate about the role of electronic correlation during the single ionization of an atomic target. This debate was fueled by several numerical models for streaked photoemission underestimating the measured relative photoemission time delay by approximately a factor of two, even though the calculations included electronic correlation at various levels of approximation~\cite{Khei10,Moor11,Nage12,Feis14,Dahl12b,Khei23}.

Accounting for electronic "shake-up" excitation during XUV photoemission from neon atoms, Isinger {\em et al.}~\cite{Isin17} compared many-body perturbation theory calculations with measured energy-integrated sideband oscillations in interferometric photoemission spectra. Based on these RABITT (reconstruction of attosecond beating by interference of two-photon transitions) experiments, the authors suggested that the discrepancy between the relative streaking time delay of $21\pm 5$~as obtained experimentally in Ref.~\cite{Schu10} and in previous streaking calculations~\cite{Khei10,Moor11,Nage12,Feis14} might be due to shake-up processes. The authors of Ref.~\cite{Isin17} worded this suggestion very carefully. They pointed out that two of their co-authors~\cite{Dahl12b} could not explain the mismatch of streaking experiments and streaking calculations by numerically modeling RABITT spectra, employing a similar theoretical many-body-perturbation theory approach as Ref.~\cite{Isin17} that includes electronic correlation. The authors provide good evidence, based on RABITT spectra, for the importance at the attosecond timescale of shake-up excitation during the photoemission of neon. However to the best of our knowledge and despite early theoretical testimony for the significance of dynamical electronic correlation on streaked photoemission~\cite{Pazo12}, the direct theoretical confirmation through streaking calculations of the $21\pm 5$~as delay in Ref.~\cite{Schu10} is still outstanding.

While the attosecond correlated ionization dynamics for multi-electron targets remains a subject of debate, remarkably, for prototypical helium, Ossiander {\em et al.}~\cite{Ossi17} found measured and calculated relative streaking time delays to agree at the (one) attosecond timescale. In their investigation, ground-state helium is singly ionized, and the dynamical response of the residual \HePlus ion is analyzed by comparing direct ionization, resulting in ground-state \HePlus ($n=1$) and shake-up ionization to a coherent superposition of \HePlus ($n = 2, 3$) sub-levels 
\begin{equation}\label{eq:t_s}
t_\text{s} = t_\text{EWS} + t_\text{CLC} + t_\text{DLC}~.
\end{equation}
Calculating $t_\text{EWS}$ and $t_\text{CLC}$ with attosecond accuracy, the authors were able to isolate a temporal shift due to electronic correlation during $n=2$ and $n=3$ shake-up ionization of about six and several ten attoseconds, respectively.

The Eisenbud-Wigner-Smith (EWS) contribution, $t_\text{EWS}$, represents the PE phase accumulation during the ionization of helium atoms in the XUV-pump pulse and includes the short-range part of the electron-electron interaction. It was calculated by direct numerical solution of the bound-continuum transition-matrix element of the two-electron system [cf., Eq.~\Eqref{amp} below]~\cite{Pazo12} using the exterior complex scaling (ECS) method~\cite{McCu04}.
While this numerical approach is state-of-the-art for helium, its value for transparently revealing physical details of the correlated dynamics during single ionization of helium is limited.

The delay contributions $t_\text{CLC}$ and $t_\text{DLC}$ are due the simultaneous interaction of the PE with the residual target charge distribution and IR-laser electric field and termed "Coulomb-laser coupling (CLC)" and "dipole-laser coupling (DLC)". 
The CLC term includes the combined effect of the monopole term of the residual charge distribution and IR-laser~\cite{Zhan10}. 
The DLC term represents the effect of the PE interaction with the dipole-component of the charge distribution of the residual \HePlus ion.
This term tends to be very small or negligible for direct ionization, but provides an important contribution for shake-up ionization, especially for large PE-detection angles relative to the XUV- and IR-pulse-polarization direction~\cite{Ossi17}. 
Applying this scheme to more complex atoms, 
such as experimentally convenient neon and argon gases~\cite{Feis14,Schu10}, is computationally challenging and requires concessions (most noticeably mean-field approximations), the accuracy of which (at the one attosecond time scale) still needs to be confirmed.

Understanding the physics behind the individual delay contributions in Eq.~\Eqref{t_s} for prototypical helium targets is crucial for the development and scrutiny of dependable attosecond streaking models for many-electron targets. It requires a highly differential study of streaked photoemission, going beyond the detection of PEs within a small solid angle about the IR- and XUV-pulse-polarization direction, in addition to the reliable modeling of the target electronic response to the incident XUV and IR pulses. 
For interferometric photoemission from helium, PE-emission-angle-dependent spectra were measured by Heuser {\em et al.}~\cite{Heus16}. These RABITT experiments detected a striking phase shift of the sideband yields for electron-detection angles that are larger than $\approx$ 50 degrees relative to the polarization direction, even in the direct emission channel (leaving \HePlus in the ground state). 
A related RABITT study for helium targets was recently carried out by Fuchs {\em et al.}~\cite{Fuch20}, who measured the three-dimensional PE-momentum distribution in coincidence with the residual ions, in order to examine experimentally the dependence of RABITT time delays (i.e., relative phase shifts between sidebands in RABITT spectra) on the PE angular momentum and coupling of continuum states in the assisting IR-laser pulse. 
Their measured delays of up to 12~as between outgoing {\it s}-- and {\it d}--wave PEs, further emphasizes the need for highly differential (angle-resolved) data for the comprehensive analysis of (streaked or interferometric) time-resolved photoemission. 

Revisiting the sensitivity of relative streaking delays on the correlated electric photoemission dynamics reported in Ref.~\cite{Ossi17} for direct and shake-up streaked photoemission from helium along the polarization direction of the XUV and IR pulse, we carried out {\em ab initio} calculations for {\em angle-resolved} direct and ($n=2,3$) shake-up photoemission.
In Secs.~\ref{sub:TDSE} and ~\ref{sub:Scatt} we review our approach for 
numerically propagating the time-dependent Schrödinger equation (TDSE) for helium exposed to external electromagnetic fields and outline the calculation of two--electron states satisfying single--ionization boundary conditions. 
Sections ~\ref{sub:tXUV} and ~\ref{sub:XUV+IR} describe our numerical calculation of different contributions (EWS, CLC, and DLC) to the experimentally observable relative streaking delay, while Sec.~\ref{sub:prop} briefly discusses noteworthy elements of our computational method, for which further details are given in the Appendices. 
Section~\ref{sec:results} contains our results for and discussion of
streaked photoemission spectra and streaking delays for direct and shake-up ionization (Sec.~\ref{sub:t_s}),
the EWS delay contribution for XUV single ionization of helium (Sec.~\ref{sub:tXUV2}),
the CLC delay contribution (Sec.~\ref{sub:CLC}), 
the DLC delay contribution in relation to the dipole moment of the residual \HePlus charge distribution (Sec.~\ref{sub:DLC}), and a multipole analysis of the evolving residual \HePlus charge distribution
and its back-action on the PE. Our summary and conclusions follow in Sec.~\ref{sec:summary}.
Throughout this work we use atomic units unless indicated otherwise.

\section{Theory} \label{sec:theory}

\subsection{Coupled radial equations} \label{sub:TDSE}

We describe the single ionization of ground-state para-helium atoms by ultrashort XUV pulses with and without assisting phase-coherent delayed (or advanced) IR-laser pulses by solving the TDSE in full dimensionality,
\begin{equation}\label{eq:TDSE}
\I \pdv{t} \Psi = H\Psi = \Big[\sum_{i=1,2} (H_i + H_{Fi} ) + V_{12}\Big] \Psi~.
\end{equation}
The Hamiltonian $H$ consists of the single-electron Hamilton operators for the interaction of each electron with the atomic nucleus, $H_i$, the electronic correlation, $V_{12}$, and the electron---external-field-interaction terms, $H_{Fi}$, given by, respectively,
\begin{equation}
\begin{aligned}
H_i & = -\frac{\laplacian_i}{2} - \frac{2}{r_i}~,\,\, \\
V_{12} &= \frac{1}{\abs{\bvec r_2 - \bvec r_1}}~, \\
\end{aligned}
\end{equation}
and 
\begin{equation} \label{eq:HFI}
H_{Fi} = \leftgroup{
\mathcal{E}(t) &\cdot z_i && (\text{length gauge})\\
-\I A(t)&\cdot \pdv{z_i} && (\text{velocity gauge})~.
}
\end{equation}
The electron position vectors,
$\bvec r_i ~,~ i=1, 2$, are chosen so that the atomic nucleus coincides with the origin of our coordinate system. 
The external electromagnetic-field pulses are assumed to be linearly polarized along the quantization ($z$) axis with cosine-squared electric-field envelopes of pulse length $T$,
\begin{equation} 
\begin{aligned}
\mathcal{E}(t) &= \mathcal{E}_\text{XUV}(t-\tau) + \mathcal{E}_\text{IR}(t) \\
\mathcal{E}_j(t) &= \mathcal{E}_{j,\text{0}}\sin(\omega_j t) \cos(\pi t/T_j)^2~,~ j = \text{XUV, IR},
\end{aligned}
\end{equation}
and corresponding vector potential $A(t) = -\int_{-\infty}^t \dd{t'} \mathcal{E}(t')$. The delay (advance) of the IR-pulse center with respect to the center of the XUV pulse is designated by 
$\tau < 0 $ ($\tau >0 $).

In our applications to photoionization by linearly polarized light,
the magnetic quantum number of the spinless two-electron system, $M=0$, is preserved and will be omitted below. The relevant set of quantum numbers needed to specify a partial wave, $\lambda = \{l_1,l_2,L\}$, thus comprises the individual angular-momentum quantum numbers $l_i$ according to $H_i~,~i=1, 2$, and the total angular-momentum quantum number $L$.
Expanding the two-electron wave function
\begin{equation} \label{eq:twoelwf}
\Psi(\bvec r_1, \bvec r_2, t) = 
{\mathcal S} \frac{1}{r_1r_2}\sum_\lambda \psi_\lambda(r_1, r_2, t) \mathcal{Y}_\lambda(\uvec r_1, \uvec r_2)
\end{equation}
in terms of real-valued generalized spherical harmonics
\begin{equation} \label{eq:genharmonics}
\begin{aligned}
&\mathcal{Y}_\lambda(\uvec r_1, \uvec r_2) =\\
&\qquad \sum_{m_1} \bmat{l_1 & l_2 & L\\ m_1 & -m_1 & 0} Y_{l_1, m_1}(\uvec r_1) Y_{l_2, -m_1} (\uvec r_2)~,
\end{aligned}
\end{equation}
we express the TDSE \Eqref{TDSE} as the set of coupled equations for the radial partial waves $\psi_{\lambda}$~\cite{Liu15}
\begin{equation} \label{eq:H_couple}
\sum_{\lambda'} H_{\lambda,\lambda'} \psi_{\lambda'} = \I\pdv{t} \psi_\lambda~,
\end{equation}
with $(r_1,r_2)$-dependent angular-coupling-matrix elements
\begin{equation}\label{eq:TDSEcouple}
H_{\lambda,\lambda'} = \mel{\mathcal{Y}_\lambda}{H}{\mathcal{Y}_{\lambda'}}~.
\end{equation}
The square brackets in Eq.~\Eqref{genharmonics} denote  Clebsch–Gordan coefficients and ${\mathcal S}$ the symmetrization operator for the two-electron spatial wave function of para-helium. Details about the matrix elements $H_{\lambda,\lambda'}$ are given in Appendix~\ref{app:TDSE}.

\subsection{Partial-wave analysis of entangled PE states} \label{sub:Scatt}

The exact PE states for single ionization of He, $\Psi$, are solutions of the time-independent Schrödinger equation,
\begin{equation}
\begin{aligned}
&H_0 \Psi = E \Psi~,\\
&H_0 = H_1 + H_2 + V_{12}~, 
\end{aligned}
\end{equation}
with energy eigenvalue $E = -Z^2/n_1^2 + k_2^2/2$. They are subject to incoming-wave boundary conditions, since the time-reversed ionization process starts with asymptotic states of specific kinetic energy, momentum, or angular momentum. Depending on the chosen boundary condition, $\Psi$ can be expressed in different ways, e.g., as states with a given total angular momentum $L$, $\mathcal S\ket{n_1,\lambda,k_2}$, or as states, $\mathcal S\ket{n_1,l_1,m_1,\bvec k_2}$, that asymptotically merge into product states. We expand the total-angular-momentum states in the basis of generalized spherical harmonics [Eq.~(\ref{eq:genharmonics})],
\begin{equation}\label{eq:EL}
\begin{aligned}
&\quad\braket{\bvec r_1, \bvec r_2}{n_1,\lambda,k_2}\\
&= \frac{1}{r_1 r_2}\sum_{l'_1,l'_2}\psi_{\lambda',n_1,k_2}(r_1, r_2)\mathcal{Y}_{\lambda'}(\uvec r_1, \uvec r_2)~,
\end{aligned}
\end{equation}
where $\lambda' = \{l_1',l_2',L\}$. The radial partial waves then satisfy the boundary condition~\cite{Bransden}
\begin{equation}\label{eq:bc}
\begin{aligned}
&\psi_{\lambda',n_1,k_2}(r_1, r_2) \overset{r_2\to\infty}{\longrightarrow} \, \delta_{l_1,l'_1}\delta_{l_2,l'_2} r_1 R_{n_1,l_1}^{(Z=2)}(r_1) \sqrt{\frac{2}{\pi}}\\
&\sin
\qty[k_2 r_2 - \frac{\pi l_2}{2} +\frac{1}{k_2}\ln(2k_2 r_2) + \sigma_{l_2} + \delta_{n_1}^{\lambda}(k_2)]~. 
\end{aligned}
\end{equation}
They depend on the radial wave function, $R_{n,l}^{(Z=2)}$,  of the residual \HePlus ion in the bound state $\psi_{n,l,m}^{(Z=2)}(\bvec r)$ and the energy-dependent Coulomb phase shift,
\begin{equation}
\sigma_l(\eta) = \arg[\Gamma(l+1+\I\eta)]~,
\end{equation}
with Sommerfeld parameter $\eta = -Z/k$. The $\sin$ factor in Eq.~\Eqref{bc} derives from the asymptotic form of Coulomb function of the first kind, $F_l(kr)$, and includes the additional energy-dependent phase $\delta_{n}^{\lambda}(k)$ due to the short-range electronic interaction during the emission process.

To provide the $\ket{l_2,m_2}$ partial-wave contributions of emitted electrons with asymptotic momentum magnitude $k_2$, we decouple the outgoing spherical-wave part of $\ket{n_1,\lambda,k_2}$ in Eq.~\Eqref{EL} with the unitary transformation
\begin{equation}\label{eq:sph_basis}
\begin{aligned}
&\ket{n_1,l_1,m_1,l_2,k_2} = \ket{n_1,l_1,m_1,l_2,-m_1,k_2}\\
&= \I \sum_{L} \bmat{l_1& l_2& L\\ m_1& -m_1 & 0} \E^{-\I \delta_{n_1}^{\lambda}} \ket{n_1,\lambda,k_2}~, 
\end{aligned}
\end{equation}
where the first equation indicates that we will drop the redundant quantum number $m_2=-m_1$ from now on.
This allows us to compose entangled two-electron states with asymptotic momentum vector $\bvec k_2$ by combining these partial waves into asymptotic plane-wave states, 
\begin{equation}\label{eq:plane}
\begin{aligned}
&\quad\ket{n_1, l_1, m_1, \bvec k_2}\\
&= \sum_{l_2}\frac{\I^{l_2}}{k_2} \E^{-\I \sigma_{l_2}} Y_{l_2,-m_1}^* (\uvec k_2)\ket{n_1,l_1,m_1,l_2,k_2}~, 
\end{aligned}
\end{equation}
where $l_1$ relates to the bound electron and $l_2$ to the PE. To simplify the notation, 
we drop subscripts and symmetrization operator from now on and refer to the symmetrized scattering states as
\begin{equation}
\begin{aligned}
&\ket{n, l, m, \bvec k} = \mathcal S\ket{n_1, l_1, m_1, \bvec k_2}~.
\end{aligned}
\end{equation}

\subsection{XUV only delays} \label{sub:tXUV}

Exposure of helium to just the XUV-pump pulse releases PE wave packets whose center is displaced relative to a fictitious free electron that starts moving at the atomic nucleus
with the constant final PE velocity at the instant when the center of the XUV pulse coincides with the atomic center. This "XUV-pulse-only" photoemission delay,
\begin{equation} \label{eq:t_xuv}
t_\text{XUV}(\bvec k) = t_\text{EWS}(\bvec k) + t_\text{C}(k,r)~,
\end{equation}
consists of the EWS delay $t_\text{EWS}$ and the Coulomb delay 
\begin{equation}\label{eq:t_C_def}
t_\text{C}(k, r) = \frac{1}{k^3}[1-\ln(2kr)]~.
\end{equation}
The EWS delay corresponds to the phase accumulation during the absorption of XUV radiation and release of the PE~\cite{Wign55, Carv02}. 
The Coulomb delay accounts for ongoing phase accumulations of the PE in the long-range $-1/r$ Coulomb potential of the residual \HePlus ion. It is identical for all atoms and does not converge as the PE propagates to the detector.

\subsubsection{Dipole transitions}

In first-order time-dependent perturbation theory, the photoemission yield and PE-phase information are calculated based on dipole-transition-matrix elements~\cite{ThummChap15}. For photoemission by an XUV pulse $\mathcal{E}_\text{XUV}(t)$ from the initial helium ground state, $\ket{i}$, into asymptotic spherical waves~\Eqref{sph_basis} and plane waves with momentum $\bvec k$~\Eqref{plane}, 
the dipole-matrix elements are, respectively,
\begin{equation}\label{eq:amp}
\begin{aligned}
M_\text{s} &= \mel{n, l_1, m, l_2, k}{(z_1+z_2)}{i}~,\\ 
M_p &= \mel{n,l,m,\bvec k}{(z_1 + z_2)}{i}
\end{aligned}
\end{equation}
and depend on the energy-dependent phase $\delta_n^{\lambda}$. 

For single-photon ionization from the helium ground state, dipole selection rules restrict accessible final-state quantum numbers to $L=1$, $M=m_1+m_2=0$ and odd $l_1+l_2$. Thus, for direct ionization  the only accessible final configuration is $(n,l_1,m,l_2)=(1,0,0,1)$. In contrast, for $n=2$ shake-up ionization, the five channels 
$(2,0,0,1)$, $(2,1,0,0)$, 
$(2,1,0,2)$, $(2,1,\pm 1,2)$ 
can be reached. 
Similarly, $n=3$ shake-up ionization allows the 13 final configurations
$(3,0,0,1)$, $(3,1,0,0)$, $(3,1,0,2)$, $(3,1,\pm 1,2)$,
$(3,2,0,1)$, $(3,2,\pm 1,1)$,
$(3,2,0,3)$, $(3,2,\pm 1,3)$, $(3,2,\pm 2,3)$.
In general, the number of dipole-accessible final channels is $2n(n-1)+1$.

\subsubsection{EWS delay}

The EWS delay carries element-specific information on the electronic correlation dynamics during the photo-release. It can be interpreted as the group delay experienced by the PE wave packet relative to a hypothetical reference wave packet~\cite{ThummChap15,Isin17,Carv02}. 
It excludes the long-range Coulomb delay $t_\text{C}$ and can be calculated numerically as the energy derivative of the phase of the dipole-matrix element for entangled spherical-partial-wave and plane-wave final PE states, respectively,
\begin{equation} \label{eq:tEWS}
\begin{aligned}
t_\text{EWS}^{n,l_1,m,l_2}(k) &= \pdv{E}\arg M_\text{s}~, \\
t_\text{EWS}^{n,l,m}(\bvec k) &= \pdv{E} \arg M_p~.
\end{aligned}
\end{equation}
As a measure for the net delay for direct or shake-up ionization to a specific \HePlus sub-shell, we define the weighted average of Eq.~\Eqref{tEWS}:
\begin{equation} \label{eq:tEWS_n1l1} 
t_\text{EWS}^{n,l}(\bvec k) = \frac{\sum_{m}
\abs{M_p}^2 t_\text{EWS}^{n,l,m}(\bvec k)}{\sum_{m}
\abs{M_p}^2}~.
\end{equation}
Similarly, we average over $\{l,m\}$ to define $t_\text{EWS}^n(\bvec k)$.

Instead of calculating matrix elements that yield the delays in first-order perturbation theory [cf., Eq.~\Eqref{tEWS}], we can retrieve {\it absolute} delays for ionization by the XUV-pulse directly from the (non-perturbative) two-electron wave function \Eqref{twoelwf}, based on the numerically calculated radial partial-wave functions $\psi_\lambda(r_1, r_2,t)$. This approach is particularly well suited for our applications in Sec.~\ref{sec:results}, which are based on the numerical propagation of partial-wave functions (cf., Sec.~\ref{sub:prop}). 
Projection of the numerically propagated wave function \Eqref{twoelwf} on residual hydrogenic \HePlus states $\phi_{n,l,m}^{(Z=2)}(\bvec r)$,
provides the PE probability density for direct ($n=1$) and shake-up ($n>1$) ionization 
\begin{equation} \label{eq:P_nlm}
\begin{aligned}
&\quad P_{n,l,m}(\bvec r,t)\\
&= 2\abs{\int \phi_{n,l,m}^{(Z=2)}(\bvec r') \Psi(\bvec r', \bvec r, t) (r')^2\dd{\Omega'}\dd{r'}}^2~,
\end{aligned}
\end{equation}
where the factor of 2 is due to symmetrization of $\Psi(\bvec r_1, \bvec r_2, t)$.

Assuming PEs are detected within the detector solid angle $\Omega_\text{det}$ centered around the direction $\uvec k$, the mean PE distances from the atomic nucleus are now given by integrating $\Omega$ over $\Omega_\text{det}$ as
\begin{equation} \label{eq:r_c_nl}
r_c^{n,l}(\bvec k, t) = \frac{\sum_m\int_{\Omega_\text{det}} r P_{n,l,m}(\bvec r,t) r^2\dd{\Omega}\dd{r}}{\sum_m \int_{\Omega_\text{det}} P_{n,l}(\bvec r,t)r^2 \dd{\Omega}\dd{r}}~.
\end{equation}
Similarly, we average over $\{l,m\}$ to define 
\begin{equation} \label{eq:r_c_n}
r_c^{n}(\bvec k, t) = \frac{\sum_{l,m}\int_{\Omega_\text{det}} r P_{n,l,m}(\bvec r,t) r^2\dd{\Omega}\dd{r}}{\sum_{l,m} \int_{\Omega_\text{det}} P_{n,l}(\bvec r,t)r^2 \dd{\Omega}\dd{r}}~.
\end{equation}
Omitting the integration over $\Omega$ corresponds to an infinitely high PE-detection resolution. 
Fitting the averaged radial PE displacements at times after which
\begin{itemize}
\item the PE wave packet has separated from the charge distribution of the residual ion and 
\item the PE velocity $\dot r_{c}^\alpha(\bvec k, t)$ has converged to the asymptotic velocity $\dot r_{c}^\alpha(\bvec k, t_\infty)$ within a preset tolerance 
\end{itemize}
according to 
\begin{equation} \label{eq:EWS_xuv}
\begin{aligned}
r_c^\alpha(\bvec k,t) 
&= \dot r_{c}^\alpha(\bvec k,t_\infty) 
[(t-\tau) - (t_\text{EWS}^\alpha(\bvec k) + t_\text{C}(k_f, r_c))]~, \\
\alpha &= \{n,l,m\}, \{n,l\}, n ~,
\end{aligned}
\end{equation}
we find the EWS delays $t_\text{EWS}^\alpha$ relative to the center of the XUV-pulse electric field.

\subsection{Streaked photoemission delay} \label{sub:XUV+IR}

\subsubsection{Absolute and relative streaking time delays}

The delay $t_\text{XUV}$ is not directly measurable with current experimental techniques, but contributes to the observable photoemission delay in streaked PE spectroscopy, where a probe pulse with an adjustable delay relative to the XUV-pump pulse is introduced~\cite{Hent01,ThummChap15}. The probe pulse, usually in the IR spectral range, maps temporal shifts in the photoemission process to observable shifts in asymptotic PE momenta. Recorded as a function of the pump-probe delay $\tau$, streaked PE spectra allow the extraction of {\it relative} streaking time delays $\Delta t_\text{s}$ by comparing the temporal offset between energetically separable oscillatory photoemission yields (streaking traces). This method for deriving relative streaked photoemission time delays has been applied to electron emission from different atoms~\cite{Liao15}, different initial electronic levels in the same target~\cite{Schu10}, and to identify temporal shifts between direct and shake-up photoemission~\cite{Ossi17}. Differences of absolute streaking delays are experimental observables.

{\em Absolute} streaking time delays $t_\text{s}$ for streaked photoemission from a specific electronic state relate temporal photoemission shifts to the phase of the probe-pulse electric field $\mathcal{E}_\text{IR}(t)$. Numerical modeling allows the calculation of absolute delays in two distinct schemes, as either "streaked spectral delays (SSDs)" or "PE wave-packet delays (PWDs)". We refer to SSDs as delays that are extracted from streaked PE spectra 
by fitting the center of momentum of a given streaking trace 
\begin{equation}\label{eq:fit}
p_f(\tau) = -aA_\text{IR}(\tau + t_\text{s}) + b \qquad (a,b > 0)
\end{equation}
to the IR vector potential $A_\text{IR}=-\int_{-\infty}^t \dd{t'} \mathcal{E}_\text{IR}(t')$. In contrast, we name delays that are extracted from the wave-packet displacement on a numerical grid using Eq.~\Eqref{EWS_xuv} as PWDs. The accuracy of the SSD $t_\text{s}$ relies only on the accuracy of the XUV--IR-delay-dependent PE-momentum spectrum and does not require the calculation of phase derivatives. For peak intensities between $10^{8}$-$10^{12}~\mathrm{W}/\mathrm{cm}^2$, streaking delays and their numerical fitting accuracy are largely independent of the IR-probe-pulse intensity~\cite{Zhan10,Pazo15}.

According to Eq.~\Eqref{t_s}, the absolute SSD $t_\text{s}$ is affected by $\tau$-dependent momentum shifts induced by the probe pulse, in particular the Coulomb-laser and dipole-laser shifts, $t_\text{CLC}$ and $t_\text{DLC}$, in addition to the probe-independent shift $t_\text{EWS}$.
For direct photoemission the short-range PWD contribution $t_\text{EWS}$ to $t_\text{XUV}$ in Eq.~\Eqref{t_xuv} merges into $t_\text{s}$ when a sufficiently weak delayed IR probe pulse is added. In contrast, the long-range Coulomb delay $t_\text{C}$, being sensitive to the phase accumulation along the entire PE trajectory, is affected by the probe-laser field and transformed into the Coulomb-laser-coupling delay $t_\text{CLC}$~\cite{Zhan10, Nage11}. 
We have verified that $t_\text{CLC}$ depends on the PE-energy and -detection direction. For emission along the XUV- and IR-pulse-polarization direction it is identical for direct and shake-up ionization.

\subsubsection{Entangled residual Stark states}

For shake-up ionization the probe-laser pulse causes an additional  dipole-laser-coupling temporal shift $t_\text{DLC}$~\cite{Pazo12}, even if the probe-pulse intensity is kept sufficiently low to make the quasi-static Stark effect in the ground-state-helium  entrance channel negligible (as is the case in typical streaking experiments). This delay contribution has no counterpart in $t_\text{XUV}$. In the exit channel the probe-laser mixes degenerate and nearly degenerate states of the residual ion, causing the formation of quasi-static \HePlus Stark hybrid states with dipole moment $\bvec d^{n}({\bvec k})$ as linear combinations of excited \HePlus states with opposite parity. 
The formation of the residual dipole affects the energy of the two-electron system in the probe-laser electric field and, in turn, the final PE-momentum and streaking delay~\cite{Bagg10,Pazo12,Pazo15}.   

To describe the formation of quasi-static residual Stark states during shake-up ionization, we unitarily transform the scattering states $\ket{n,l,m,\bvec k}$~\Eqref{plane} to  Stark scattering states with the same ($n, m$) quantum numbers and total energy $-Z^2/n^2 + k^2/2$,
\begin{equation} \label{eq:Starkstate}
\ket{n,\alpha,m,\bvec k} 
= \sum_l C_{\alpha,l}^{n,m} \ket{n,l,m,\bvec k}~,
\end{equation}
where $\alpha$ denotes the Stark quantum numbers~\cite{Bransden}. The expansion coefficients $\{C_{\alpha,l}^{n,m}\}$ for the $n=2,3$ shells of \HePlus are listed in Tabs.~\ref{tab:dipole_n2} to \ref{tab:dipole_n3_m1} of Appendix \ref{app:Stark}.

After shake-up ionization by the XUV pulse, once the PE wave packet and residual \HePlus charge distribution effectively no longer overlap, we represent 
the residual \HePlus ion in a given 
$n$ channel as a superposition of degenerate Stark states, 
\begin{equation}  \label{eq:bound_Psi}
\ket*{\psi_\text{\HePlus}^{(n)}(\bvec k)} = \mathcal N\sum_{\alpha,m} c_{n,\alpha,m}(\bvec k)\ket{n,\alpha,m}~,
\end{equation}
with normalization constant $\mathcal{N} = (\sum_{\alpha,m} \abs{c_{n,\alpha,m}}^2)^{-1/2}$. To obtain the expansion coefficients from the numerically  provided accurate solution $\Psi$ of the two-electron TDSE~\Eqref{TDSE} for XUV-only ionization, we project at a time $t^\star = 5.5$ fs after the center of XUV pulse
onto the two-electron entangled scattering states $\bra{n,\alpha,m,\bvec k}$ in Eq.~(\ref{eq:Starkstate}),
\begin{equation}  \label{eq:c2s}
c_{n,\alpha,m}(\bvec k) = 
\braket{n,\alpha,m,\bvec k}{\Psi(t^\star)}~.
\end{equation}
For photoemission along the XUV-pulse-polarization direction ($\uvec k = \uvec z$), only 
$m=0$ terms contribute in Eq.~\Eqref{bound_Psi}. Thus,   
$\ket*{\psi_\text{\HePlus}^{(n)}(\bvec k)}$ is rotationally symmetrical about the $z$ axis and its dipole moment,
\begin{equation}\label{eq:vec_dipole}
\bvec d^{(n)}({\bvec k}) = -\mel*{\psi_\text{\HePlus}^{(n)}(\bvec k)}{\bvec r}{\psi_\text{\HePlus}^{(n)}(\bvec k)}~,
\end{equation}
aligned with $\uvec k$. In contrast, for PE-detection directions $\uvec k$ that are not aligned with $\uvec z$, the combination of terms with $m\neq0$ results in electric dipole vectors \Eqref{vec_dipole} that, generally, do not align with $\uvec z$.

A sufficiently weak, quasi-static IR-streaking electric field does not noticeably distort the electronic structure of the initial He atom and residual ion. Its interaction with the XUV-pulse-generated dipole yields the energy shift
\begin{equation}\label{eq:E-dip}
\begin{aligned}
\Delta E_\text{dip}^{(n)}({\bvec k})  &= {\mathcal E}_\text{IR}(\tau) 
\mel*{\psi_\text{\HePlus}^{(n)}(\bvec k)} z {\psi_\text{\HePlus}^{(n)}(\bvec k)}  \\
&= - \dv{A_\text{IR}}{\tau} \;\;  d^{(n)}_z({\bvec k})
\end{aligned}
\end{equation}
and corresponding PE momentum shift
\begin{equation}\label{eq:dk2_sub}
\begin{aligned}
\Delta k^{(n)} &\approx \frac{-\Delta E_\text{dip}^{(n)}({\bvec k})}{k}
= -\dv{A_\text{IR}}{\tau} \frac{d^{(n)}_z({\bvec k})}{k}~.
\end{aligned}
\end{equation}
This equation builds on the assumption that for shake-up emission to a specific residual shell $n$, PE wave packets corresponding to different asymptotic states
$\{ \ket{n,\alpha,m} \}$ in Eq.~\Eqref{bound_Psi} have the same center of momentum 
$k = \sqrt{2(\omega_\text{XUV} - I^{(n)}_p)}$.  
Our numerical results in Sec.~\ref{sub:DLC} below validate this assumption by revealing that, for PE momenta $k$ within the spectral range of the XUV pulse, the relative magnitudes of the coefficients $c_{n,\alpha,m}(\bvec k)$  keep their proportions (cf. \Figref{sis2n}).

At large PE -- \HePlus distances, the PE has accumulated the EWS phase shift, 
and the PE-momentum shift in the oscillating IR field can be approximated as~\cite{Pazo15},
\begin{equation}\label{eq:k_f}
\bvec k_f(\tau) \approx \big(k + \Delta k^{(n)}\big) \uvec k
-\bvec A_\text{IR}(\tau + t_\text{EWS})~.
\end{equation}
Denoting the polar angle of $\bvec k_f$ and $\bvec k$ as $\theta_f$ and $\theta$, respectively, and defining $\gamma = \theta_f-\theta$, projection of Eq.~\Eqref{k_f} onto $\uvec k_f$  yields
\begin{equation}\label{eq:k_f_appr}\begin{aligned}
k_f(\tau) &= k\cos\gamma \\
&- \cos\theta_f\Big[A_\text{IR}(\tau + t_\text{EWS}) + t_\text{DLC}\dv{A_\text{IR}(\tau + t_\text{EWS})}{\tau}\Big]\\
&\approx k\cos\gamma - \cos\theta_f \cdot A_\text{IR}(\tau + t_\text{EWS} + t_\text{DLC})~,
\end{aligned}\end{equation}
where
\begin{equation}\label{eq:tDLC}
t_\text{DLC}(\bvec k) = \frac{\cos\gamma}{\cos\theta_f}\frac{d_z^{(n)}(\bvec k)}{k}
\approx \frac{d_z^{(n)}(\bvec k)}{k \cos\theta}~.
\end{equation}
For the IR-pulse peak intensity considered in this work, $\gamma$ is less than $1.6^\circ$, making it negligible in Eq.~\Eqref{tDLC}. The approximation in Eq.~\Eqref{k_f_appr} is supported by our typical delays $t_\text{DLC}$ being less than 5\% of the IR period. 
We note that an alternative approximation is to treat $A_\text{IR}(\tau)$ as sinusoidal~\cite{Ossi17}, which results in
\begin{equation}\label{eq:tDLC2}
t_\text{DLC}(\bvec k) \approx \frac{1}{\omega_\text{IR}}\tan^{-1}\qty[\omega_\text{IR} \frac{d_z^{(n)}(\bvec k)}{k\cos\theta}]~.
\end{equation}

\subsection{Propagation algorithm} \label{sub:prop}
 
We solve the TDSE \Eqref{TDSE} using a finite-element discrete-variable representation (FE-DVR) of the partial-wave radial wave functions $\psi_\lambda(r_1, r_2, t)$ on a 2D adaptive rectangular numerical grid for the radial coordinates $(r_1, r_2)$. For details of our implementation of the FE-DVR method we refer to Ref.~\cite{Liu14}, references therein, and Appendices \ref{app:TDSE} and \ref{app:optim} of the present work.

\subsubsection{Length gauge}

Expressing the electron interactions with the XUV and IR pulse \Eqref{HFI} in the length gauge, we time-propagate the radial TDSE \Eqref{H_couple} using the split-operator Crank-Nicolson technique~\cite{Band94} with equidistant time steps $\Delta t$, 
\begin{equation}
\begin{aligned}
&\psi_{\lambda}(t+\Delta t) = \E^{-\I H\Delta t} \psi_{\lambda}(t)\\
&\quad = \E^{-\I H_0\frac{\Delta t}{2}}\E^{-\I H_\text{int}\Delta t}
\E^{-\I H_0\frac{\Delta t}{2}}\psi_{\lambda}(t) + \order{\Delta t^3} ~.
\end{aligned}
\end{equation}
This scheme assumes the time-dependence of $H$ to be negligible during each time step. In all applications in Sec.~\ref{sec:results} below, we ascertain numerical convergence by selecting sufficiently small time steps for the accumulation of numerical errors in each time step being irrelevant.

The advantageous choice of $H_0 = H_1 + H_2$ allows the factorization of the propagation along the radial coordinates,
\begin{equation}
\E^{-\I H_0\frac{\Delta t}{2}} = 
\E^{-\I H_1\frac{\Delta t}{2}}\E^{-\I H_2\frac{\Delta t}{2}},
\end{equation}
since $H_1$ and $H_2$ commute. 
In the length gauge, the interaction operator $H_\text{int} = H_{F1}+H_{F2}+V_{12}$ in position representation does not contain derivatives. It thus couples all partial waves separately at any given point of the 2D numerical grid (without the need for error-susceptible finite differencing). The split-operator method is therefore particularly suitable for length-gauge calculation.
For all exponentiations we use the Cayley form of the time-evolution operator~\cite{NR3},
\begin{equation}
\E^{-\I H\Delta t} = \qty(1+\frac{\I}{2} H \Delta t)^{-1} \qty(1-\frac{\I}{2}  H \Delta t)~.
\end{equation}

\subsubsection{Velocity gauge}

For applications in Sec.~\ref{sec:results} below we employ both,  the length and velocity form of $H_{Fi}$ [cf., Eq.~\Eqref{HFI}], in order to assess the numerical convergence of our PE-momentum distribution and temporal shifts. While formally quantum mechanics is gauge invariant, approximations introduced in the modeling and the progression of numerical errors tend to affect the accuracy of length- and velocity-gauge calculations unequally. Their comparison is therefore a common tool for assessing the accuracy of numerical models. In particular, as pointed out by Cormier and Lambropoulos~\cite{Corm96} for the numerical computation of photoemission spectra for strong-field ionization based on partial-wave expansions, the rate of convergence in length and velocity-gauge calculations with the number of included partial waves tends to be different. For above-threshold ionization of hydrogen atoms, the authors obtained converged spectra including 10 partial waves in velocity-gauge calculations, while requiring in excess of 200 partial waves in the length gauge. The physical reason for the very different rate of convergence is a more effective inclusion of the external-field-driven electronic quiver-motion in velocity-gauge final PE states in term of the canonical PE momentum, $\bvec k - \bvec A(t)$. In contrast, the same anisotropic dynamics requires a large number of partial waves in the length gauge, where the final-state description hinges on the (time-independent) kinematic PE momentum $\bvec k$.

While this convergence comparison appears to clearly favor the velocity gauge, the split-operator scheme outlined above in part offsets its advantageous inclusion of the canonical PE momentum due to the appearance of the differential operator $\pdv*{z_i}$ in Eq.~\Eqref{HFI}. In contrast to the length gauge and represented in spherical coordinates, the electron--external-field interaction $H_{Fi}$ now involves the coupling of two-electron partial waves at adjacent grid points that are addressed in the finite-difference implementation of the $\pdv*{z_i}$ operator on the ($r_1,r_2$) grid. For this reason, we abandon the split-operator propagation scheme when working in the velocity gauge and, instead, use a Lanczos algorithm~\cite{Pazo12} with an adaptive number of Krylov basis states to keep the numerical error below a set threshold. 

In a Krylov basis $\{\Psi, H\Psi, \ldots, H^{K-1} \Psi \}$ with $K$ states, where $\Psi$ is the wave function at the current propagation time, $H$ in Eq.~\Eqref{TDSE} is represented by a tridiagonal matrix
\begin{equation}
\mat H =
\begin{pmatrix}
\alpha_0 & \beta_1 &&& \\ 
\beta_1 & \alpha_1 & \beta_2 && \\ 
 & \beta_2 & \alpha_2 & \ddots& \\ 
& & \ddots& \ddots & \beta_{K-1} \\
&&&\beta_{K-1} & \alpha_{K-1}
\end{pmatrix}~.
\end{equation}
The truncation error in the Lanczos algorithm we use to diagonalize $\mat H$ can be estimated as~\cite{McCu04}
\begin{equation}
\epsilon_\text{lanc}^{K} \approx \frac{\Delta t^K}{K!} \prod_{i=1}^K \beta_i~.
\end{equation}
We dynamically adjust $K$ at each propagation time step $\Delta t$ to satisfy $\epsilon_\text{lanc}^{K-1} < 10^{-12}$, rather than using a fixed number of basis states as, e.g., Ref.~\cite{Ossi17}.

\section{Results and analysis} \label{sec:results}

\subsection{Streaking delays} \label{sub:t_s}

\begin{figure}[h] 
\centering{}\includegraphics[width=1\linewidth]{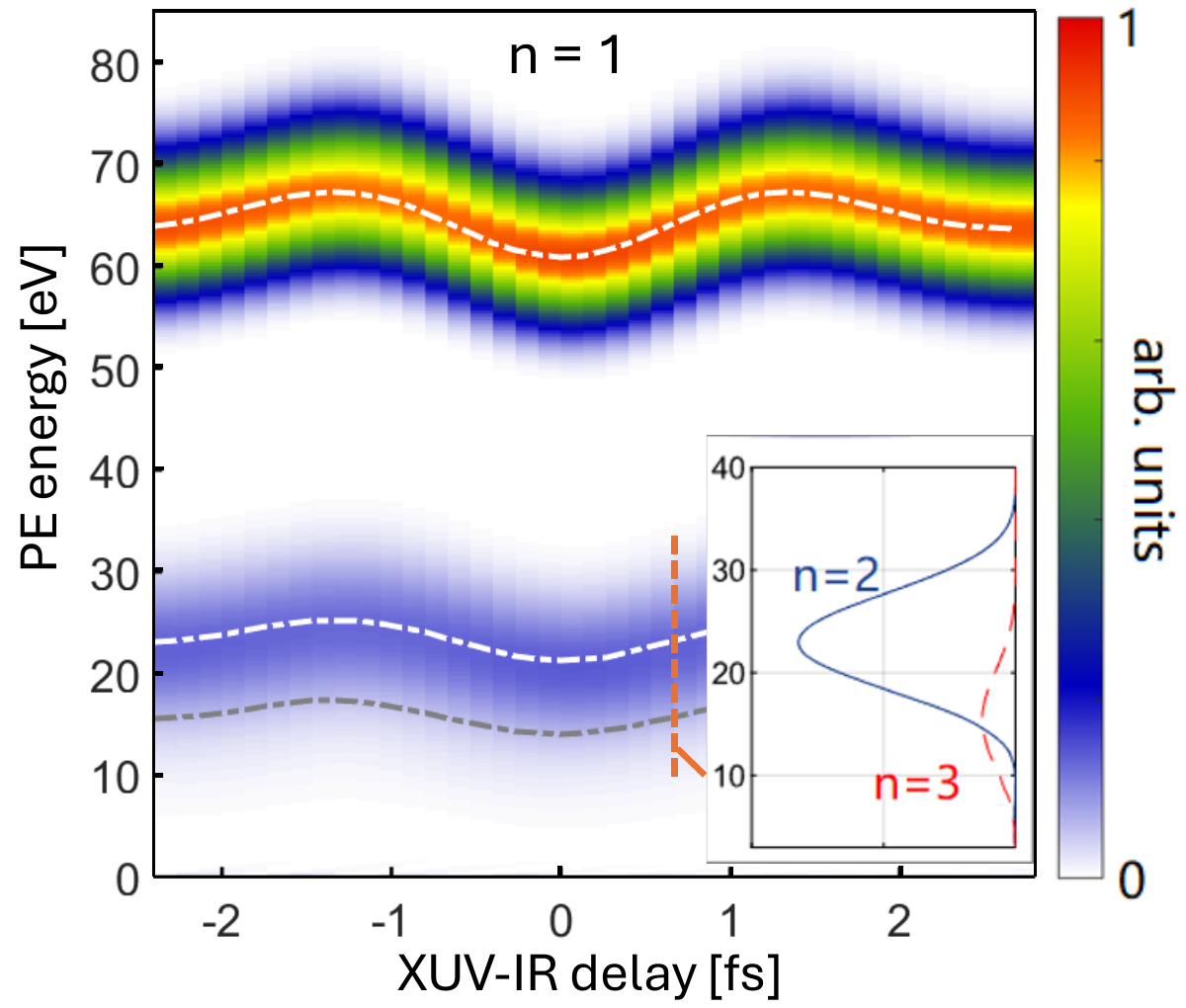}
\caption{Streaked photoemission spectra for ground-state para-helium for a photon energy of 90~eV XUV. The centers of energy for direct ($n=1$) and $n=2$ shake-up emission are shown as white dash-dotted lines. The gray dash-dotted line is the center of energy for $n=3$ shake-up emission. The inset shows line-outs at 
$\tau= T_{IR}/4 = 0.6671$~fs of the $n=2$ and $n=3$ shake-up emission yields.}
\label{fig:streaking}
\end{figure}

Following~\cite{Ossi17}, we assume an IR streaking pulse with a Full-Width Half-Intensity Maximum (FWHIM) of 3~fs and a peak intensity of $4\times10^{11}\ \mathrm{W/cm^2}$. The XUV pulse has a photon energy of 90~eV, FWHIM of 200~as, and peak intensity of $10^{12}\ \mathrm{W/cm^2}$. 
We use 40 equally spaced XUV--IR-pulse delays for a delay range of 2 IR periods, calculate the center of PE momenta within an infinitely narrow detection cone, and fit to the IR-pulse vector potential to extract the streaking delays according to Eq.~\Eqref{fit}. Figure \ref{fig:streaking} shows our corresponding calculated streaked photoemission spectrum. The streaking traces for direct and $n=2$ shake-up emission are clearly separated.

\begin{figure}[h]
\centering{}\includegraphics[width=1\linewidth]{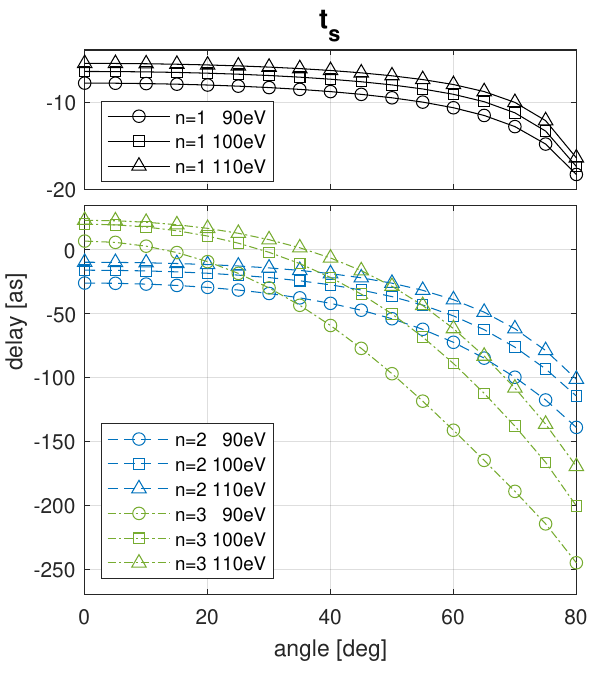}
\caption{Electron-emission-angle-dependent absolute direct ($n=1$) and shake-up ($n=2$) streaking delays $t_\text{s}$ for XUV photon energies of 90, 100, and 110~eV. 
}
\label{fig:ang_delay}
\end{figure}

Absolute streaking delays $t_\text{s}$ for direct and $n=2,3$ shake-up emission extracted from the spectrum in \Figref{streaking} are show in \Figref{ang_delay}, together with delays for XUV photon energies of 100 and 110~eV. The dependence of $t_\text{s}$ on the XUV-pulse photon energy and PE-detection angle for emission is complex, owing to the three distinct terms in Eq.~\Eqref{t_s} and different \HePlus $(n,l,m)$ sub-channels. We therefore separately examine contributions to $t_\text{s}$ in the following subsections. The most prominent feature, a rapid decrease of $t_\text{s}$ with PE-detection angle, is primarily due to the $1/\cos\theta$ factor in the DLC term [cf. Eqs.~\Eqref{t_s} and \Eqref{tDLC}]. 

\subsection{XUV-only delays} \label{sub:tXUV2}

\begin{figure}[h]
\centering{}\includegraphics[width=1\linewidth]{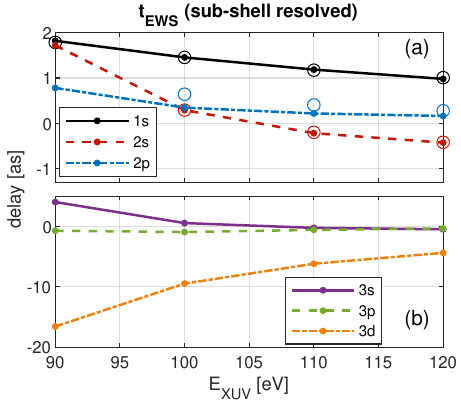}
\caption{Absolute $(n,l)$ sub-channel-resolved EWS delays $t_\text{EWS}$ for emission along the XUV-polarization direction and XUV-photon energies of 90, 100, 110, and 120~eV. (a) Direct (black line), \HePlus (2$s$) (red dashed line), and \HePlus (2$p$) (blue dash-dotted line) shake-up emission. Open circles show corresponding numerical results of Pazourek {\em et al.}~\cite{Pazo12} 
(b) 3$s$ (purple line), 3$p$ (green dashed line), and 3$d$ (orange dash-dotted line) shake-up emission.
}
\label{fig:tEWS}
\end{figure}

Figure \ref{fig:tEWS} shows the EWS time delay, $t_\text{EWS}$, extracted from the PE wave packet according to Eqs.~\Eqref{r_c_nl} and \Eqref{EWS_xuv}, for direct ionization ($n=1$) and all $(n,l)$ shake-up sub-channels of the ionic L and M shells. These results are calculated for emission along the XUV-pulse polarization and XUV photon energies of 90, 100, 110, and 120~eV. Lines in between the small markers are added to guide the eye. For direct (black line) and 2$s$ shake-up (red dashed line) our results in \Figref{tEWS} (a) agree with the phase-derivative calculation of Pazourek {\em et al.}~\cite{Pazo12} according to Eqs.~\Eqref{tEWS} and \Eqref{tEWS_n1l1}. For shake-up emission to \HePlus (2$p$) our results still agree with Ref.~\cite{Pazo12} with attosecond accuracy and show small discrepancies only at the sub-attosecond level. Generally, in the absence of resonances, EWS delays tend to decrease in absolute value for increasing XUV photon energy, due to the decreasing time left for strong PE interactions. This trend is followed for direct, 2$p$-, and 3$p$-shake-up emission. At the sub-attosecond time scale,  the onset for this expected decrease is not reached for 2$s$-shake-up emission for XUV photon energy up to 120~eV.  

\begin{figure}[h] 
\centering{}\includegraphics[width=1\linewidth]{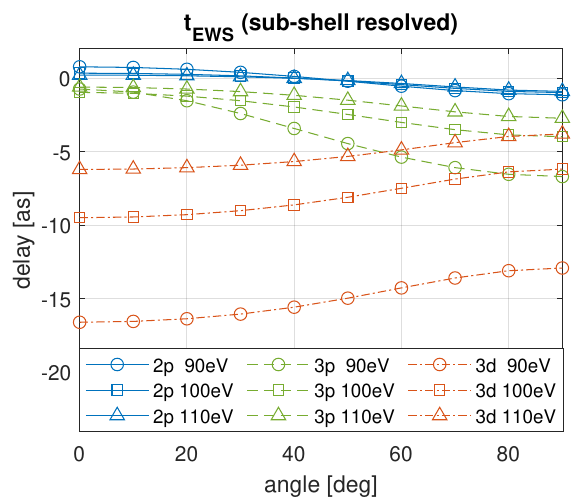}
\caption{Electron-emission-angle-dependent absolute EWS delays $t_\text{EWS}$ for XUV photon energies of 90, 100, and 110~eV and specific \HePlus $(n,l)$ shake-up sub-shells. 
}
\label{fig:tEWS_sub}
\end{figure}

Figure~\ref{fig:tEWS_sub} extends the sub-shell-resolved EWS delays, shown in \Figref{tEWS}  for emission along the XUV-polarization axis, to arbitrary PE-detection directions for XUV photon energies of 90, 100, and 110~eV. 
For 3$p$ and 3$d$ shake-up emission this graph clearly shows the expected decrease in magnitude of the EWS delay for increasing XUV photon energy and a more pronounced dependence on the PE-detection direction in comparison to 2$p$ shape-up emission. Noticeable are also the general increase of the delay magnitude with the principle and angular momentum quantum numbers of the residual shake-up \HePlus shake-up state, in compliance with the larger extent of the residual charge distributions. 
Delays for 1$s$, 2$s$, 3$s$ shake-up emission are angle independent, shown in \Figref{tEWS}, and not duplicated in this graph.

\begin{figure}[h]
\centering{}\includegraphics[width=1\linewidth]{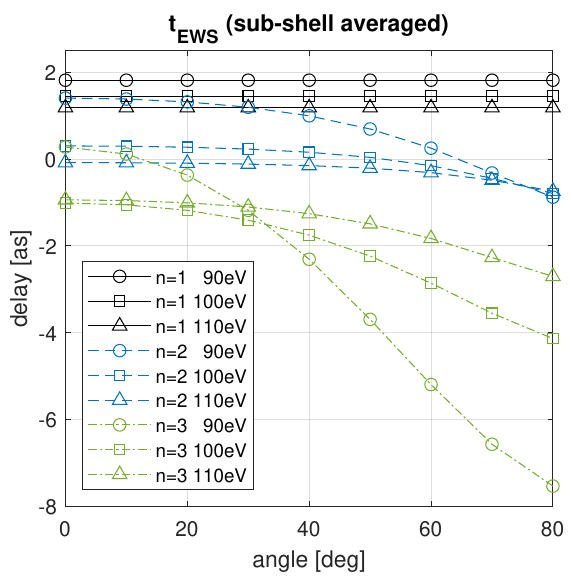}
\caption{Electron-emission-angle-dependent absolute EWS delays $t_\text{EWS}$ for direct ($n=1$) and $n=2,3$ shake-up emission for XUV photon energies of 90, 100, and 110~eV. The shake-up delays are averaged over $(l,m)$ sub shells according to Eq.~\Eqref{r_c_n}. 
} 
\label{fig:tEWS_ang}
\end{figure}

Averaging the EWS photoemission delays for $n=2$ and $3$ according to Eqs.~\Eqref{r_c_n} and \Eqref{EWS_xuv} results in the emission-angle-dependent absolute sub-shell averaged delays displayed in \Figref{tEWS_ang}, in addition to the angle-independent delays for direct XUV photoemission.
The angular dependence is most pronounced for shake-up emission at the lowest XUV photon energy (90~eV). In comparison with the sub-shell-resolved delays in \Figref{tEWS_sub}, the angular dependence of the sub-shell-averaged delays in~\Figref{tEWS_ang}
tends to follow more closely the 3$p$ than the 3$d$ delays. 
The less uniform angle dependence of the sub-shell-averaged delays for different photon energies, in particular, the crossing of $n=3$ delays in \Figref{tEWS_ang}, 
is a result of the sub-shell and XUV-photon-energy- and angle-dependent emission probabilities $P_{n,l,m}(\bvec r_2,t)$ employed as weights in Eq.~\Eqref{r_c_n}.

To keep the contribution of the EWS delay to the absolute XUV delay \Eqref{t_xuv} in perspective, we note that for direct emission by 90 eV XUV pulses the non-convergent Coulomb delay amounts to $t_\text{C} = -53.8,-57.0,-59.1\ \text{(as)}$ at detector distances $r_2=0.5,2,5\ \text{(m)}$, respectively. The negative long-range Coulomb delay thus constitutes an advance that dominates the EWS delay.

\subsection{The Coulomb-laser-coupling delay} \label{sub:CLC}

\subsubsection{Definition of CLC delays for hydrogen atoms}

The CLC delay accounts for the long-range PE-phase accumulation. For emission along the XUV- and IR-pulse-polarization directions it does not noticeably depend on the target electronic structure and is approximately identical for H and He~\cite{Pazo12}. It is largely a classical quantity, which was found to agree within $\sim 1$ as for quantum and classical-trajectory Monte Carlo calculations~\cite{Pazo15}. We assume its long-range classical behavior to apply for any final PE-momentum direction $\uvec k$ (i.e, for any PE detection angle $\theta$) and calculate it {\em ab initio},  following Refs.~\cite{Pazo12,Pazo15}, according to
\begin{equation}\label{eq:HCLC}
t_\text{CLC}(\bvec k) = t^{\text{H}_{100}}_\text{s}(\bvec k) - t^{\text{H}_{100}}_\text{EWS}(k)~,
\end{equation}
based on the  streaking [$t^{\text{H}_{100}}_\text{s}(\bvec k)$] and EWS delay 
\begin{equation}\label{eq:HEWS}
\begin{aligned}
t^{\text{H}_{100}}_\text{EWS}(k) &= \pdv{E} \arg \mel{C(\bvec k)}{z}{1,0,0} \\
&= \eval{\pdv{\sigma_{l=1}}{E}}_{E=k^2/2}~
\end{aligned}\end{equation}
for ground-state hydrogen with outgoing Coulomb waves 
\begin{equation}\begin{aligned}
&\braket{\bvec r}{C(\bvec k)} =\\
&\qquad \frac{1}{r}\sqrt{\frac{2}{\pi}}\sum_{l,m} \qty[\frac{\I^l}{k} \E^{-\I\sigma_l}Y_{l,m}^* (\uvec k)] F_l(\eta, kr) Y_{l,m}(\uvec r)~.
\end{aligned}\end{equation}
We note that $t^{\text{H}_{100}}_\text{EWS}$ is angle independent due to the symmetry of the hydrogen ground state. 

\begin{figure}[h]
\centering{}\includegraphics[width=1\linewidth]{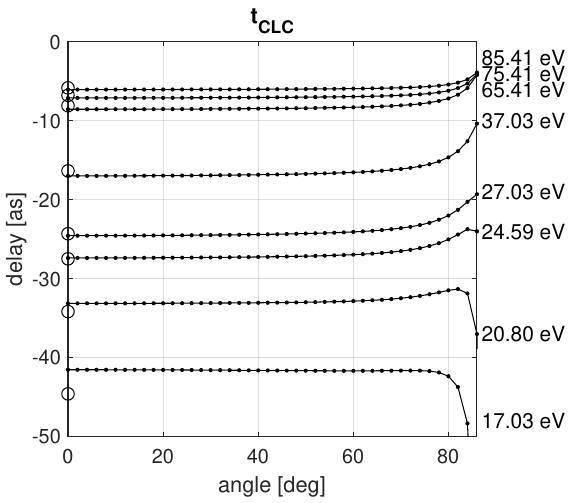}
\caption{CLC delays for hydrogen as functions of the PE-detection direction for 
final PE energies  between 17.03 and 85.41~eV (dots). Circles show delays calculated in Ref.~\cite{Pazo15} for photoemission along the XUV- and IR-pulse-polarization direction. 
}
\label{fig:CLC}
\end{figure}

Figure~\ref{fig:CLC} shows the CLC delay for hydrogen as a function of the PE-detection angle for final PE energies between 17.03 and 85.41~eV, as given by Eq.~\Eqref{HCLC}. As expected, $\abs{t_\text{CLC}}$ decreases with increasing XUV photon energy (or asymptotic PE kinetic energy). 

\subsubsection{Classical-trajectory CLC calculation}

To investigate the angular dependence for PE emission to large angles in \Figref{CLC}, we extended the classical calculation for emission along the XUV-pulse-polarization direction in Ref.~\cite{Pazo15} and performed an angle-resolved classical-trajectory simulation. We assume instant absorption of a single XUV photon with energy $E_\text{XUV}$ to provide PEs at aligned initial positions and momenta, $\bvec r_0 \propto \uvec{z} $ and $\bvec k_0 \propto \uvec{z}$, respectively, with total energy
\begin{equation}
E_\text{PE} = E_\text{XUV} + E_{100}^{(Z)} = \frac{\bvec k_0^2}{2} - \frac{Z}{r_0}~,
\end{equation}
where $E_{100}^{(Z)}$ is the ground-state energy of a hydrogen-like atom. This equation gives $k_0$ as a function of $r_0$. We determine $r_0 = 0.55$ a.u., such that the classically calculated CLC delay, $t_\text{CLC}^\text{cl}$, matches the results in \Figref{CLC} at zero degrees, and apply this value for all energies considered in this figure. 

Propagating classical PE trajectories starting at $(\bvec r_0, \bvec k_0)$,  subject to the same IR field, we use for He streaking calculation for a range of XUV--IR pulse delays $\tau$, we obtain $\tau$-dependent asymptotic PE momenta, i.e., classical streaking traces, that allow us to extract the classical streaking delay $t_\text{s}^{\text{H}_{100},\text{cl}}$. Substitution into Eq.~\Eqref{HCLC} now yields delays $t_\text{CLC}^\text{cl}$ that are {\em independent} of the PE detection direction, suggesting that the angle dependence of $t_\text{CLC}$ in \Figref{CLC} is a quantum-mechanical effect.

\subsection{The dipole-laser-coupling delay} \label{sub:DLC}

Exposure of the residual dipole \Eqref{vec_dipole} to the electric field of the IR streaking pulse results in the energy shift
$  
\Delta E_\text{dip} = - d_z^{(n)}({\bvec k}) \, {\mathcal E}_\text{IR}(\tau)
$  
in Eq.~\Eqref{E-dip}.
This shift affects the entangled PE--residual-ion system and results in the DLC contribution $t_\text{DLC}$ to the streaking delay in Eq.~\Eqref{t_s}.
The effect of $t_\text{DLC}$ for PE emission along the XUV-polarization direction is discussed in Ref.~\cite{Pazo12}. In this subsection we examine $t_\text{DLC}(\bvec k)$ for arbitrary PE detection directions $\uvec k$.

\begin{figure}[h]
\centering{}\includegraphics[width=0.8\linewidth]{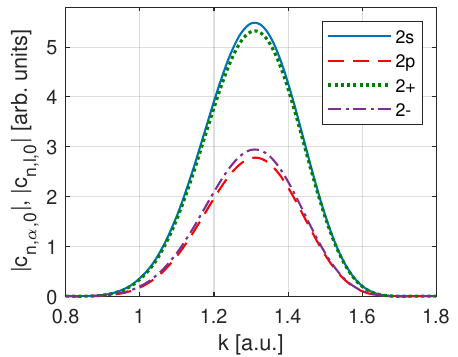}
\caption{Absolute values of the probability amplitudes of scattering states for $n=2$ shake-up emission from helium as a function of the PE momentum magnitude $k$. The probability amplitudes are represented in the Stark basis according to Eq.~\Eqref{c2s}. Labels "2$^\pm$" indicate Stark states with quantum numbers $\alpha = \pm 1$ of the residual \HePlus ion. Hydrogenic states are labeled as "2$s$" and "2$p$".}
\label{fig:sis2n}
\end{figure}

Applying Eq.~\Eqref{tDLC}, we calculate the coefficients $c_{n,\alpha,m}(\bvec k)$ according to Eq.~\Eqref{c2s}.
For the XUV-pulse parameters considered in the present work, we find that the ratios of the magnitudes of the complex-valued  coefficients $c_{n,\alpha,m}(\bvec k)$ remain approximately constant over spectral profile of the PE wave packet. This allows us to evaluate these coefficients at the central PE momentum $k = \sqrt{2(\omega_\text{XUV} - I^{(n)}_p)}$. Figure \ref{fig:sis2n} illustrates this proportionate change for the example of $n=2$ shake-up emission. 

\begin{figure}[h]
\centering{}\includegraphics[width=1\linewidth]{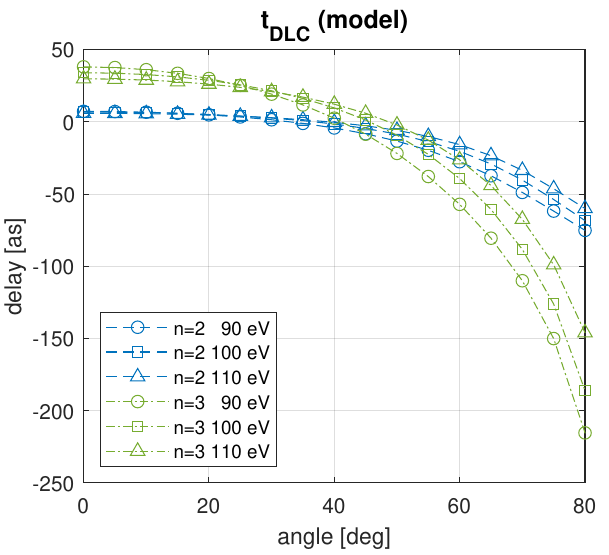}
\caption{Model DLC delays according to Eq.~\Eqref{tDLC} for $n=2$ and $n=3$ shake-up ionization of helium and XUV pulses with photon energies of 90, 100, and 110~eV 
as functions of the PE-detection direction. 
For a comparison with results for 
emission along the XUV-polarization direction of Ossiander {\em et al.}~\cite{Ossi17} see Tab.~\ref{tab:DLC_compare}.
}
\label{fig:tDLC_model}
\end{figure}

The approximate DLC delays $t_\text{DLC}$ for $n=2$ and $n=3$ shake-up ionization of para-helium and XUV pulses with photon energies of 90, 100, and 110~eV show a striking PE-detection-direction dependence at large angles that is easily traced to the $1/\cos\theta$ factor in the dipole model given by Eq.~\Eqref{tDLC} (\Figref{tDLC_model}). 
Within this model, direct photoemission has no DLC delay, since $d_z^{(0)}=0$.
The sign change for shake-up emission at $\theta \approx 40^\circ$ indicates a change in the orientation of the residual dipole at larger angles. 

\begin{table}[H]
\centering
\caption{Comparison of our {\em ab initio} calculated and  modeled DLC delays for emission along the XUV- and IR-pulse-polarization direction
with the correlation delays of Ossiander {\em et al.} \cite{Ossi17} derived from measured streaking delays. This table complements \Figref{tDLC_model}.} \label{tab:DLC_compare}
\vspace{0.3cm} 
\begin{tabular}{|c|c|c|c|c|c|}
\hline
$n$  & $\hbar \omega_\text{XUV}$ & {\em ab initio} & DLC model & Ossiander {\em et al.} \cite{Ossi17} \\
     & $ [eV]$ & [as] & [as] & [as] \\
\hline \hline
$2$ &   $90$  & 1.4 & 7.3 & 7.8 \\
\hline
$2$ &  $100$  & 2.5 & 6.7 & 6.55 \\
\hline
$2$ &  $110$  & 3.9 & 5.9 & 5.6 \\
\hline \hline
$3$ &  $90$   & 48.4 & 38.1 & 45.9 \\
\hline
$3$ &  $100$  & 44.8 & 34.0 & 37.7 \\
\hline
$3$ &  $110$  & 40.2 & 29.8 & 32.2 \\
\hline
\end{tabular}
\end{table}

\begin{figure}[h]
\centering{}\includegraphics[width=1\linewidth]{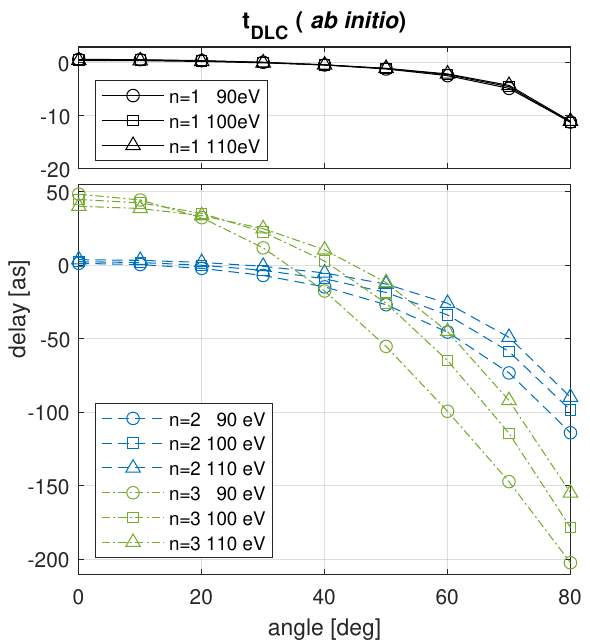}
\caption{{\em Ab initio} angle- and energy-resolved absolute DLC delays 
for direct ($n=1$) and $n=2, 3$ shake-up ionization of helium for XUV-pulse photon energies of 90, 100, and 110~eV as functions of the PE-detection direction.
}
\label{fig:tDLC}
\end{figure}

We validate the DLC model results in \Figref{tDLC_model} by computing streaking, EWS, and CLC delays {\em ab initio}. The {\em ab initio} DLC delays shown in \Figref{tDLC} are obtained from Eqs.~\Eqref{t_s} and \Eqref{HCLC} as 
\begin{equation} 
 t_\text{DLC}(\bvec k) = t_\text{s}(\bvec k)  - t^{\text{H}_{100}}_\text{s}(\bvec k) 
    -[t_\text{EWS}(\bvec k) - t^{\text{H}_{100}}_\text{EWS}(\bvec k)] ~,
\end{equation}
in good overall agreement with the model DLC delays for shake-up emission in \Figref{tDLC_model}. For most angles, this agreement is better than 10 as. It becomes worse at large angles near $\theta = 90^\circ$, possibly due to a combination of the model assumptions leading to Eq.~\Eqref{tDLC} and 
small numerical signal-to-noise ratios (very small yields) at large detection angles. As seen in Tab.~\ref{tab:DLC_compare}, for the special case of PE emission along the XUV-polarization direction (0$^\circ$), our DLC-model delays are in good quantitative agreement with the correlation delays 
derived from measured streaking delays in Fig.~S10 of the supplementary material of Ref.~{\em et al.}~\cite{Ossi17}.

The DLC model does not explain the angle dependence of the {\em ab initio} DLC delay for direct emission shown in the upper graph of \Figref{tDLC}. In addition, at the as timescale, it is at variance with our {\em ab initio} calculated delays (Tab.~\ref{tab:DLC_compare}). 
A possible cause for this variation is that the approximate calculation of $t_\text{CLC}$ for hydrogen according to Eq.~\Eqref{HCLC} does not accurately represent the CLC delay for helium. 
On the other hand, accurately calculating the CLC delay for direct photoemission from helium as 
\begin{equation}\label{eq:HeCLC}
t_\text{CLC}^\text{He}(\bvec k) = t^{n=1}_\text{s}(\bvec k) - t_\text{EWS}^{n=1}(k)~.
\end{equation}
for the $n=1$ curves in \Figref{tDLC} would merely illustrate the difference of the two definitions, $t_\text{CLC}^\text{He} - t_\text{CLC}$, instead of a deviation from the  physically intuitive DLC model \Eqref{tDLC}.

\subsection{Residual multipole analysis} \label{sub:multipo}

During photoemission, the electronic probability density of the residual \HePlus ion keeps evolving while interacting with the PE. After a sufficiently long time after the photo-release by the XUV pulse, the PE wave-function overlap with the charge distribution of the residual electron is small enough for the conditional electronic probability density,
\begin{equation}
P^{\text{He}^{+}}(\bvec r, t) = \frac{\abs{\Psi(\bvec r, \bvec r_\text{PE} ,t)}^2}{\int \abs{\Psi(\bvec r', \bvec r_\text{PE}, t)}^2 \dd[3]{r'}}~,
\end{equation}
to be associated with the residual ion's electronic charge density, where the PE is assumed as a classical particle located at the maximum of the PE wave-packet probability, $\bvec r_\text{PE}$.

The multipole expansion of the electronic potential energy $V^{\text{He}^{+}}$ corresponding to the charge density $P^{\text{He}^{+}}$, assumed to be confirmed within a volume of radius $a < r_\text{PE}$, results in
\begin{equation} \label{eq:multipole}
V^{\text{He}^{+}}(\bvec r,t) = \sum_{l=0}^\infty\sum_{m=-l}^l \frac{1}{r^{l+1}} C_{l,m}(t) Y_{l,m}(\uvec r) \quad (r > a)~,
\end{equation}
\begin{equation} \label{eq:multipole-coeff}
C_{l,m}(t) = \frac{-4\pi}{(2l+1)} \int_{r\leqslant a} r^l Y_{l,m}^*(\uvec r) P^{\text{He}^{+}}(\bvec r, t) \dd[3]{r}~. 
\end{equation}
Using individual terms of this expansion as effective potentials in separate SAE calculations allows the quantitative comparison of the effects of given multipole orders on PE spectra and streaking time delays. Performing such simplified one-electron calculations for the parameters of the present work, we found dominant monopole and dipole interactions, while quadrupole and higher multipole orders are negligible.  

\begin{figure*} 
\centering{}\includegraphics[width=1\linewidth]{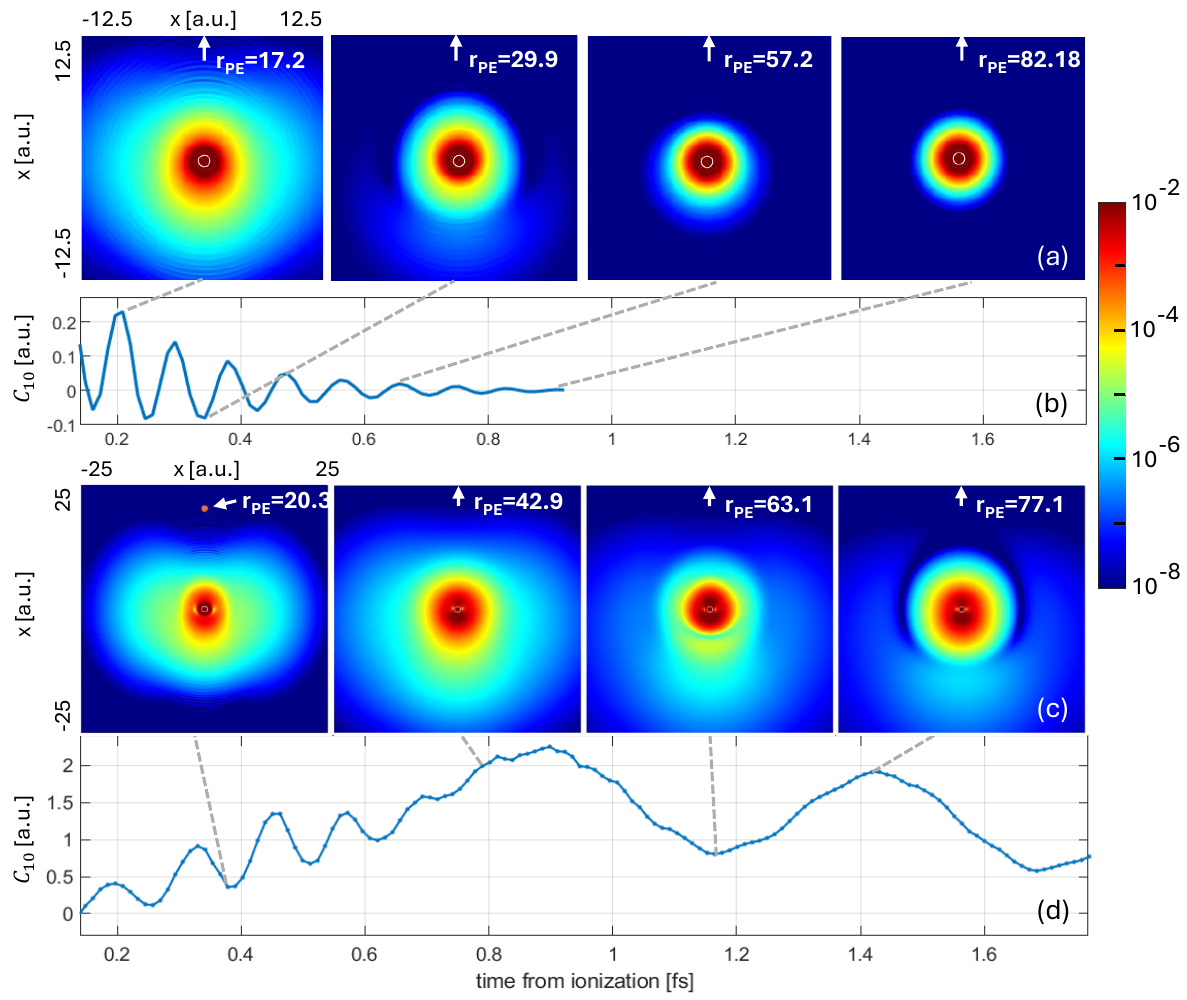}
\caption{ Evolution of the residual \HePlus electronic charge distribution after photoionization  of helium by 90~eV, $10^{12}\ \mathrm{W/cm^2}$ XUV pulses with a FWHIM of 200~as for (a) direct and (c) shake-up emission. 
(b) and (d): Corresponding dipole coefficients $C_{1,0}(t)$ in the multipole expansion of the residual-ion potential energy [Eq.~\Eqref{multipole}]. The distance from the nucleus of the (assumed classical) PE, $r_\text{PE}$, is given in a.u. and pointed to by white arrows. 
}
\label{fig:multi_pole}
\end{figure*}

A few snapshots of the residual charge-density evolution are shown in \Figref{multi_pole}
for the ionization of para-helium in 90~eV, $10^{12}\ \mathrm{W/cm^2}$ XUV pulses with a FWHIM of 200 as (no IR pulse is present). 
We assumed emission along the XUV-polarization direction (more specifically, the $+z$ axis), such that only multipole terms with $m=0$ are present. The color-coded graphs show residual charge distributions $P^{\text{He}^{+}}(\bvec r, t)$ for increasing classical PE distances $r_\text{PE}$ from the atomic nucleus. Residual probability densities for direct and shake-up emission are displayed in Figs.~\ref{fig:multi_pole} (a) and (c), respectively. The corresponding coefficients $C_{1,0}(t)$ of the dipole term in the multipole expansion \Eqref{multipole} for direct and shake-up emission are shown in Figs.~\ref{fig:multi_pole} (c) and (d), respectively. 

The oscillation period of the contribution for direct emission in \Figref{multi_pole} (b) is about 88~as, corresponding to an energy of 47~eV, close to the energy difference between the undistorted K and L shells of \HePlus, 40.8~eV. We therefore attribute these small-amplitude oscillations to weak quantum beats of the \HePlus ground state against transiently populated $n=2$ levels. The rapid decay of these beats as the PE moves away from the residual ion illustrates the noticeable presence of electronic entanglement, even for direct emission and PE distances way outside the residual charge distribution.

For shake-up ionization the composition and evolution of the residual charge cloud are more complex than for direct emission [\Figref{multi_pole} (d)]. Now the dipole term starts oscillating with a period of about 120~as (35 eV). 0.7~fs after the ionization the oscillation period changes to 530 as (7.8 eV). These two quantum-beat frequencies are close to undistorted K- to L-shell and L- to M-shell excitation energies of \HePlus, respectively. As expected, the oscillation amplitude for shake-up emission in \Figref{multi_pole} (d) is significantly larger than for direct emission in \Figref{multi_pole} (b). In addition, the oscillations for shake-up ionization persist long after those for direct emission have decayed. These findings are in agreement with an intuitively expected stronger correlation of the PE with electronically excited \HePlus.

\subsection{The polarization effect by IR field}

In addition to the XUV photoionization dynamics generating polarized residual \HePlus ions, the electric field of the IR streaking pulse   contributes to the polarization of the residual ion and may thereby affect angle-dependent shake-up photoemission yields and streaking delays. In this section we investigate this effect and show that, for the assumed IR-pulse intensity in the present work, the effect of the residual-ion polarization by the streaking pulse can be neglected.   

For hydrogen atoms this effect was investigated before by Baggesen and Madsen~\cite{Bagg10},  who pointed out that the polarization of a target atom by the streaking IR-pulse may influence the angle-dependence of streaked photoemission yields and delays. They showed that, for sufficiently strong IR streaking fields, the streaked PE asymptotic energy acquires an additional phase shift relative to 
the IR vector potential $A_\text{IR}(t)$, due to the fact that the dipole-laser coupling is proportional to the IR electric field $\mathcal{E}_\text{IR}(t)$ and thus phase-shifted by 90$^\circ$.

To quantify the distortion of the  \HePlus residual ion by the streaking field, we calculate its static polarizability~\cite{Bransden}, 
\begin{equation}\begin{aligned} \label{eq:He+Polariz}
\beta_{n, l, 0}  & = \beta_{n, l, 0}^{b} + \beta_{n, l, 0}^{c} \\
 & =2\sum_{n',l'}^{n'\ne n} \frac{\abs{\mel{n',l',0}{z}{n,l,0}}^2}{E_{n'}^{(0)}-E_n^{(0)}}  \\
 & +2\sum_{l'}\int_0^\infty \dd{k} \frac{\abs{\mel{k,l',0}{z}{n,l,0}}^2}{k^2/2-E_n^{(0)}}~,  
\end{aligned}\end{equation}
based on virtual excitations to hydrogenic bound states, $\{ \ket{n,l,m} \}$, ($\beta_{n, l, 0}^{b}$) and spherical partial-wave continuum states, $ \ket{k,l',m} $, ($\beta_{n, l, 0}^{c}$) for the nuclear charge $Z=2$. Our numerical values for the lowest four states of \HePlus are in close agreement with the scaled theoretical values in Refs.~\cite{Baye12,Drak06} (Tab.~\ref{tab:He+Pol}). The literature values in Tab.~\ref{tab:He+Pol} are calculated for hydrogen atoms and multiplied by $1/16$ to account for the scaling $\beta_{n,l,0} \propto 1/Z^4$~\cite{Bransden,Drak06}. 
\begin{table}[H]
\centering
\caption{{\em Ab initio} static polarizabilities of \HePlus in a.u. according to Eq.~\Eqref{He+Polariz} in comparison with calculations by Baye~\cite{Baye12}. $\bar{m}$ indicates values averaged over $m$ within an ($n,l$) sub-shell.} \label{tab:He+Pol}
 \vspace{0.3cm} 
\begin{tabular}{|l|c|c|c|c|}
\hline
$n,l,m$  & $\beta_{n,l m}^{b}$ & $\beta_{n,l,m}^{c}$ & $\beta_{n,l,m}$ &  $\beta_{n,l,m}$\cite{Baye12} \\
\hline \hline
$1,0,0$ &   0.2287 & 0.0520 &  0.2807 & 0.2812 \\
\hline \hline
$2,0,0$       &   6.917  & 0.571  & 7.488  & 7.500  \\
\hline
$2,1,0$       &  12.98   & 0.508  &  13.49 &        \\
\hline
$2,1,\pm 1$   & 9.391    & 0.351  & 9.742  &        \\
\hline
$2,1,\bar{m}$ & 10.59    & 0.403  & 10.99  & 11.00  \\
\hline \hline
$3,0,0$       &  60.94   & 2.231  &  63.17 & 63.28  \\
\hline
$3,1,0$       & 98.50    & 2.646  &  101.2 &        \\
\hline
$3,1,\pm 1$   & 69.08    &  1.702 &  70.78 &        \\
\hline
$3,1,\bar{m}$ & 78.89    &  2.017 &  80.91 & 81.00  \\
\hline
$3,2,0$       &  145.3   & 1.445  &  146.7 &        \\
\hline
$3,2,\pm 1$   & 130.3   & 1.248   & 131.5  &       \\
\hline
$3,2,\pm 2$   & 85.24   & 0.759   & 86.00  &        \\
\hline
$3,2,\bar{m}$ & 115.3   &  1.092  &  116.4 & 116.4  \\
\hline
\end{tabular}
\end{table}

For the IR-pulse peak intensity assumed in this work of $4\times 10^{11} \mathrm{W/cm^2}$, corresponding to a peak-electric-field strength of $\mathcal{E}_\text{IR,0} = 3.4\e{-3}$ a.u., the distortion of the electric dipole moment of the \HePlus ground state is 
\begin{equation} \label{eq:dd100}
\Delta d_{1,0,0,\text{max}} = 2\, \mathcal{E}_\text{IR,0}\, \beta_{1,0,0} = 1.9\e{-3}~, 
\end{equation}
where the factor of two derives from the maximal variation of the IR electric field over one IR oscillation period. 
For the $n=2$ shell of \HePlus, the largest dipole polarizability is $\beta_{2,1,0} = 13.49$, resulting in  
\begin{equation} \label{eq:dd_excited}
\Delta d_{2,1,0,\text{max}} = 2\mathcal{E}_\text{IR,0}\beta_{2,1,0} \approx 9.2 \e{-2}~.   
\end{equation}
The maximal IR-pulse-induced dipole in the $n=2$ shell is thus about 50 times larger than for the ground state, and the distortion of the \HePlus ground state may be neglected in comparison with the distortion of excited \HePlus states by the streaking IR field.

At the same time, the maximal IR-pulse-induced dipole in the $n=2$ shell in Eq.~(\ref{eq:dd_excited}) is about 16 times smaller than the dipole moment of 1.5 of the XUV-pulse-excited $n=2$ Stark states according to Tab.~\ref{tab:dipole_n2} in Appendix~\ref{app:Stark}. This is consistent with the numerical example 
in Fig.~\ref{fig:multi_pole}, where the maximal dipole coefficient for $n=2$ shake-up emission exceeds the maximal dipole coefficient for direct emission by an order of magnitude. The distortion by the IR streaking pulse of the XUV-photoemission-produced residual dipole is thus a small correction.  
This difference is less pronounced for $n=3$ shake-up excitation. Here, the  XUV-photoemission-produced residual-dipole moment is 4.5 (cf., Tabs.~\ref{tab:dipole_n3_m0} and \ref{tab:dipole_n3_m1} in Appendix~\ref{app:Stark}), exceeding 6.5 times the IR-pulse-induced dipole distortion,
\begin{equation}
\Delta d_{3,1,0,\text{max}} = 2\mathcal{E}_\text{IR,0}\beta_{3,1,0} \approx 0.69~.  \end{equation}
These comparisons indicate that, for shake-up ionization into the K and L shells of \HePlus, the distortion of the XUV-produced dipole by the IR pulse is negligible, while for $n=3$ shake-up ionization  further scrutiny is needed in this regard.

We note that once the IR distortion of \HePlus becomes relevant, the distinction of three separate delay contributions to the streaking delay in Eq.~(\ref{eq:t_s})  breaks down, and the EWS term needs to be modified to account for the influence of the IR-pulse during photoemission for small XUV -- IR pulse delays. In the present work,  the fact that the three contributions in Eq.~(\ref{eq:t_s})  correctly add up to the streaking delay, provides further evidence for the IR-pulse distortion of \HePlus being negligible.  
Furthermore, the classical  trajectory analysis in Sec.~\ref{sub:CLC}, excluding the IR-pulse distortion of the residual ion, agrees for emission angles below $\approx$60$^\circ$ with the quantum result and further supports the very minor role of the IR distortion of the residual ion in the present work.  

This analysis of the distortion of the residual ion by the IR pulse may provide a reference for futures streaking-pulse parameters that break the non-distorting dipole assumption. For a discussion of the IR-pulse distortion of helium in the entrance channel, we refer to Ref.~\cite{Pazo12}. For the description of relative streaking delays, the IR-pulse distortion of the neutral helium target affects direct and shake-up channels equally and thus tends to cancel in the calculation of relative streaking delays.

\section{Summary and conclusions} \label{sec:summary}

\label{subsec:summary}

In this work we confirmed the results of previous theoretical and experimental investigations for streaked direct and shake-up photoemission from helium atoms along the IR- and XUV-pulse linear polarization. We performed {\em ab initio} and model calculations and extended these earlier investigation to  arbitrary PE detection directions, scrutinizing EWS, CLC, and DLC contributions for direct and \HePlus($n=2,3$) shake-up ionization.

For large PE detection angles we found that CLC delays become angle dependent and tracked this dependence as a purely quantum-mechanical effect that cannot be explained with classical-trajectory simulations. For shake-up ionization we find dominant contributions to the streaking delay due to DLC. Confirming a physically intuitive model for the DLC effect, we validated the model predictions for the DLC delay against {\em ab initio} calculations over a large range of PE detection angles. 

Distinguishing the production of polarized residual \HePlus ions due to the XUV-photoemission dynamics and possible contributions from its time-dependent polarization by the oscillating IR pulse, we estimated the IR-pulse-induced polarization contribution. Our model DLC calculations without including the IR distortion of the XUV-pulse-generated dipole agreeing with the DLC-delay contribution derived from {\em ab initio} calculations provide evidence for the IR-pulse distortion of the residual dipole being negligible.

\appendix

\section{Coupling-matrix elements and Poisson integral} \label{app:TDSE}

Each term in the Hamiltonian of Eq.~\Eqref{TDSE} can be represented as a band-diagonal matrix in an FE-DVR basis, and the computational cost for the propagation by one time step scales linearly with the total number of grid points. 
As basis function, we use polynomials defined by Gauss-Lobatto quadrature, scaled to fit each FE interval for the FE-DVR expansion of the radial partial-wave functions,
\begin{equation}
\psi_\lambda(r_1, r_2, t) = \sum_{i,j} c_{i,j}\phi_i(r_1)\phi_j(r_2)~.
\end{equation}
These basis functions are localized on the grid points of the 2D numerical ($r_1$ and $r_2$) grid and approximately orthonormal,
\begin{equation}\label{eq:fedvr_orth}
\phi_i(x_j) \sim \delta_{ij}, \qquad \int \phi_i(x)\phi_j(x)\dd{x} \approx \delta_{ij}~,
\end{equation}
where $ \{x_j\} $ designates the set of grid points for either one of the radial coordinates, $r_1$ and $r_2$.

The numerical efficiency of the FE-DVR method derives from these polynomials being adapted to and defined on the employed numerical grid. 
The flexibility of the FE-DVR method stems from its applicability to any given (possibly unequally spaced) set of FE boundary points.

\subsection{Electron-electron interaction}

The electronic interaction term, $V_{12}$, in the Hamiltonian is expanded in generalized spherical harmonics,
\begin{equation}
 \begin{aligned}
V_{12} = 4\pi \sum_{l=0}^{\infty} \frac{(-1)^l}{\sqrt{2l + 1}} \frac{r_ < ^l}{r_ > ^{l+1}} \mathcal{Y}_{l,l}^0 (\uvec r_1, \uvec r_2)~. 
\end{aligned}
\end{equation}
For the angular part of the electronic coupling-matrix elements we have
\begin{equation}
 \begin{aligned}
 & \mel{\mathcal{Y}_\lambda}{\mathcal{Y}_{l,l}^0}{\mathcal{Y}_{\lambda'}}
= \frac{2l+1}{4\pi} \delta_{L,L'} \sqrt{(2l_1'+1)(2l_2'+1)(2L+1)} \\
& \qquad \times \begin{bmatrix}l & l_1' & l_1\\ 0 & 0 & 0\end{bmatrix} 
 \begin{bmatrix}l & l_2' & l_2\\ 0 & 0 & 0\end{bmatrix} 
 \qty{\begin{matrix}l & l_1' & l_1\\ l & l_2' & l_2\\ 0 & L & L\end{matrix}}
\end{aligned} \end{equation}
and calculate the radial matrix elements accurately using Poisson integrals~\cite{McCu04},
\begin{equation}\label{eq:poisson}
 \begin{aligned}
 &\mel{\phi_i \phi_j}{\frac{r_ < ^l}{r_ > ^{l+1}}}{\phi_{i'} \phi_{j'}} =\delta_{i,i'} \delta_{j,j'}\\
 & \qquad \times \qty[\frac{2l+1}{r_i r_j \sqrt{\omega_i \omega_j}} (2 \mat T_l)^{-1}_{i,j} + \frac{r_i^l r_j^l}{r_{max}^{2l+1}}]~,
\end{aligned}
\end{equation}
where
\begin{equation}
\begin{aligned}
(T_l)_{ij} &= \mel{\phi_i}{ \qty[-\frac{1}{2m} \frac{\mathrm{d}^{2}}{\mathrm{d}{r}^{2}} + \frac{l(l+1)}{2mr} ] }{\phi_j} \\
&= -\frac{1}{2m} \mel{\phi_i}{ \frac{\mathrm{d}^{2}}{\mathrm{d}{r}^{2}} }{\phi_j} + \delta_{i,j}\frac{l(l+1)}{2mr_i^2}~.
\end{aligned}
\end{equation}

\subsection{External-field interaction}

The length-gauge external-field-interaction term $H_{Fi}$ in the TDSE [Eq.~\Eqref{TDSE}] contributes to the matrix element in Eq.~\Eqref{H_couple} and is evaluated according to
\begin{equation}
\mel{\mathcal Y_\lambda}{\mathcal E(t)\cdot z_i}{\mathcal Y_{\lambda'}} = \mathcal E(t) r_i F_{\lambda,\lambda'}^{(i)}\quad (i=1,2)~,
\end{equation}
where the angular integrals are
\begin{equation}\label{eq:F1}
\begin{aligned}
& F_{\lambda,\lambda'}^{(i)} = \mel{\mathcal{Y}_\lambda}{ \cos\theta_i}{\mathcal{Y}_{\lambda'}} \\
&=\sqrt{9(2l'_1+1)(2l'_2+1)(2L'+1)/(4\pi)^2}  
  \begin{bmatrix}\delta_{i,1} & l'_1 & l_1\\ 0 & 0 & 0\end{bmatrix} \\
&\times 
  \begin{bmatrix}\delta_{i,2} & l'_2 & l_2\\ 0 & 0 & 0\end{bmatrix}
  \begin{bmatrix}1 & L' & L\\ 0 & 0 & 0\end{bmatrix}
   \qty{\begin{matrix} \delta_{i,1} & l'_1 & l_1\\ \delta_{i,2} & l'_2 & l_2\\ 1 & L' & L\end{matrix}} \quad (i=1,2)~,
\end{aligned}
\end{equation}
and the curly brackets denote the Wigner 9j symbols.

In the velocity gauge, $H_{Fi}$ couples the angular function as 
\begin{equation}
\mel{\mathcal Y_\lambda}{-\I A(t) \pdv{z_i}}{\mathcal Y_{\lambda'}} = -\I A(t) G_{\lambda,\lambda'}^{(i)}\quad (i=1,2)~,
\end{equation}
where
\begin{equation}\label{eq:Gi}
G^{(i)}_{\lambda,\lambda'} = F^{(i)}_{\lambda,\lambda'}\pdv{r_i} + \frac{g^{(i)}_{\lambda,\lambda'}}{r_i}~
\end{equation}
and
\begin{equation}
g_{\lambda,\lambda'}^{(i)} = -\mel{\mathcal Y_\lambda}{\cos\theta_i +\sin\theta_i\pdv{\theta_i}}{\mathcal Y_{\lambda'}}~.
\end{equation}
In particular, the hydrogenic field-interaction matrices are~\cite{Baue06})
\begin{equation}\label{eq:HeTDSE_21}\begin{aligned}
&g_{\lambda,\lambda'}^{(1)} =\\
& \delta_{l_2, l_2'} \sum_{m_1} \bmat{l_1 & l_2 & L\\ m_1 & -m_1 & 0}\bmat{l_1' & l_2' & L'\\ m_1 & -m_1 & 0} \times\\
& (\delta_{l_1',l_1+1} l'_1\mathcal C_{l_1,m_1} - \delta_{l_1,l'_1+1} l_1\mathcal C_{l'_1,m_1})
\end{aligned}\end{equation}
and 
\begin{equation}\label{eq:HeTDSE_22}\begin{aligned}
&g^{(2)}_{\lambda,\lambda'} =\\
&\delta_{l_1, l_1'} \sum_{m_2} \bmat{l_1 & l_2 & L\\ -m_2 & m_2 & 0}\bmat{l_1' & l_2' & L'\\ -m_2 & m_2 & 0} \times\\
& (\delta_{l_2',l_2+1} l_2'\mathcal C_{l_2,m_2} - \delta_{l_2,l'_2+1} l_2 \mathcal C_{l'_2,m_2})~,
\end{aligned}\end{equation}
with
\begin{equation} \label{eq:SphHar_20}
\begin{aligned}
\mathcal C_{l,m}
= \sqrt{\frac{(l+1)^2-m^2}{(2l+1)(2l+3)}}~.
\end{aligned}
\end{equation}

\section{Stark states} \label{app:Stark}

The following tables give the matrix element of the unitary transformation between the asymptotic \HePlus
Stark states $ \ket{n,\alpha,m} $ in Eq.~\Eqref{Starkstate} and $Z=2$ hydrogenic bound states $ \ket{n,l,m} $ within a given shell $n$. 
The first column  shows the eigenvalues of the electric dipole moment $d_z^{(n,\alpha,m)}$ for each Stark state.
\begin{table}[H]
\centering
\caption{Stark states for $n=2, m=0$} \label{tab:dipole_n2}
\begin{tabular}{|c||c|c|c|}
\hline
$d_z^{(n,\alpha,m)}$ & \diagbox{$\ket{n,\alpha,m}$}{$\ket{n,l,m}$} & $\ket{2,0,0}$ & $\ket{2,1,0}$ \\
\hline
$3/2$ & $\ket{2,1,0}$ & $1/\sqrt2$ & $1/\sqrt2$ \\
\hline
$-3/2$ & $\ket{2,-1,0}$ & $1/\sqrt2$ & $-1/\sqrt2$ \\
\hline
\end{tabular}
\end{table}

\begin{table}[H]
\centering
\caption{Stark states for $n=3, m=0$} \label{tab:dipole_n3_m0}
\begin{tabular}{|c||c|c|c|c|}
\hline
$d_z^{(n,\alpha,m)}$ & \diagbox{$\ket{n,\alpha,m}$}{$\ket{n,l,m}$} & $\ket{3,0,0}$ & $\ket{3,1,0}$ & $\ket{3,2,0}$\\
\hline
$9/2$ & $\ket{3,1,0}$ & $1/\sqrt3$ & $1/\sqrt2$ & $1/\sqrt6$ \\
\hline
$0$ & $\ket{3,0,0}$ & $1/\sqrt3$ & $0$ & $-\sqrt{2/3}$ \\
\hline
$-9/2$ & $\ket{3,-1,0}$ & $1/\sqrt3$ & $-1/\sqrt2$ & $1/\sqrt6$ \\
\hline
\end{tabular}
\end{table}

\begin{table}[H]
\centering
\caption{Stark states for $n=3, m=\pm1$} \label{tab:dipole_n3_m1}
\begin{tabular}{|c||c|c|c|}
\hline
$d_z^{(n,\alpha,m)}$ & \diagbox{$\ket{n,\alpha,m}$}{$\ket{n,l,m}$} & $\ket{3,1,\pm1}$ & $\ket{3,2,\pm1}$ \\
\hline
$9/4$ & $\ket{3,1,\pm1}$ & $1/\sqrt2$ & $1/\sqrt2$ \\
\hline
$-9/4$ & $\ket{3,-1,\pm1}$ & $1/\sqrt2$ & $-1/\sqrt2$ \\
\hline
\end{tabular}
\end{table}
These matrices are calculated by diagonalizing $z$ in each degenerate subspace spanned by all states $\{ \ket{n,l,m} \}$ for fixed $n,m$~\cite{Bransden}. The subspaces with $n=2, m=\pm 1$ and $n=3, m=\pm 2$ are non-degenerate and have zero polarization with Stark quantum number $\alpha=0$. The $z$ component of the \HePlus dipole in Eqs.~(\ref{eq:vec_dipole}) and (\ref{eq:E-dip}) is calculated as
\begin{equation}
\begin{aligned}
d^{(n)}_z({\bvec k}) &= -\mel*{\psi_\text{\HePlus}^{(n)}(\bvec k)}{z}{\psi_\text{\HePlus}^{(n)}(\bvec k)}~,\\
&= -{\mathcal N}^2 \sum_{\alpha,m} \abs{c_{n,\alpha,m}(\bvec k)}^2 d_z^{(n,\alpha,m)}~,
\end{aligned}
\end{equation}
where the coefficients $c_{n,\alpha,m}(\bvec k)$ are defined in Eq.~\Eqref{c2s}.

\section{Optimizations} \label{app:optim}

\subsection{Selection rules} 

For para-helium $\Psi(\bvec r_1, \bvec r_2)$ is symmetric with respect to electron exchange. Spherical harmonics satisfying the relation
\begin{equation}\label{eq:GenYlm_5}
\mathcal{Y}_\lambda(\uvec r_1, \uvec r_2) = 
(-1)^{l_1+l_2-L} \mathcal{Y}_\lambda(\uvec r_2, \uvec r_1)~,
\end{equation}
implies 
\begin{equation}\label{eq:HeTDSE_1}
\psi_{l_1, l_2}^L(r_1, r_2,t) = (-1)^{l_1 + l_2 - L} \psi_{l_2, l_1}^L(r_2, r_1,t)
\end{equation}
and eliminates about half of the partial waves in Eq.~\Eqref{twoelwf}, allowing us to keep only those for $l_1 \leqslant l_2$. This simplification is compromised for the rectangular (non-quadratic) grids we used, where $r_1$ and $r_2$ have different spacial ranges, and $l_1$ and $l_2$ require different maximal values for numerical convergence.  

The symmetry properties of the Clebsch-Gordan  and Wigner 9j coefficients imply that  Eq.~\Eqref{TDSEcouple} couples partial waves with the same odd values of $l_1+l_2-L$. Since the initial ground state of para-helium ($L=0$, $l_1=l_2$) has even $l_1+l_2-L$, partial waves  with odd $l_1+l_2-L$ do not contribute.

\subsection{Adaptive grid with optional sliding window}

\begin{figure}[h]
\centering{}\includegraphics[width=0.61\linewidth]{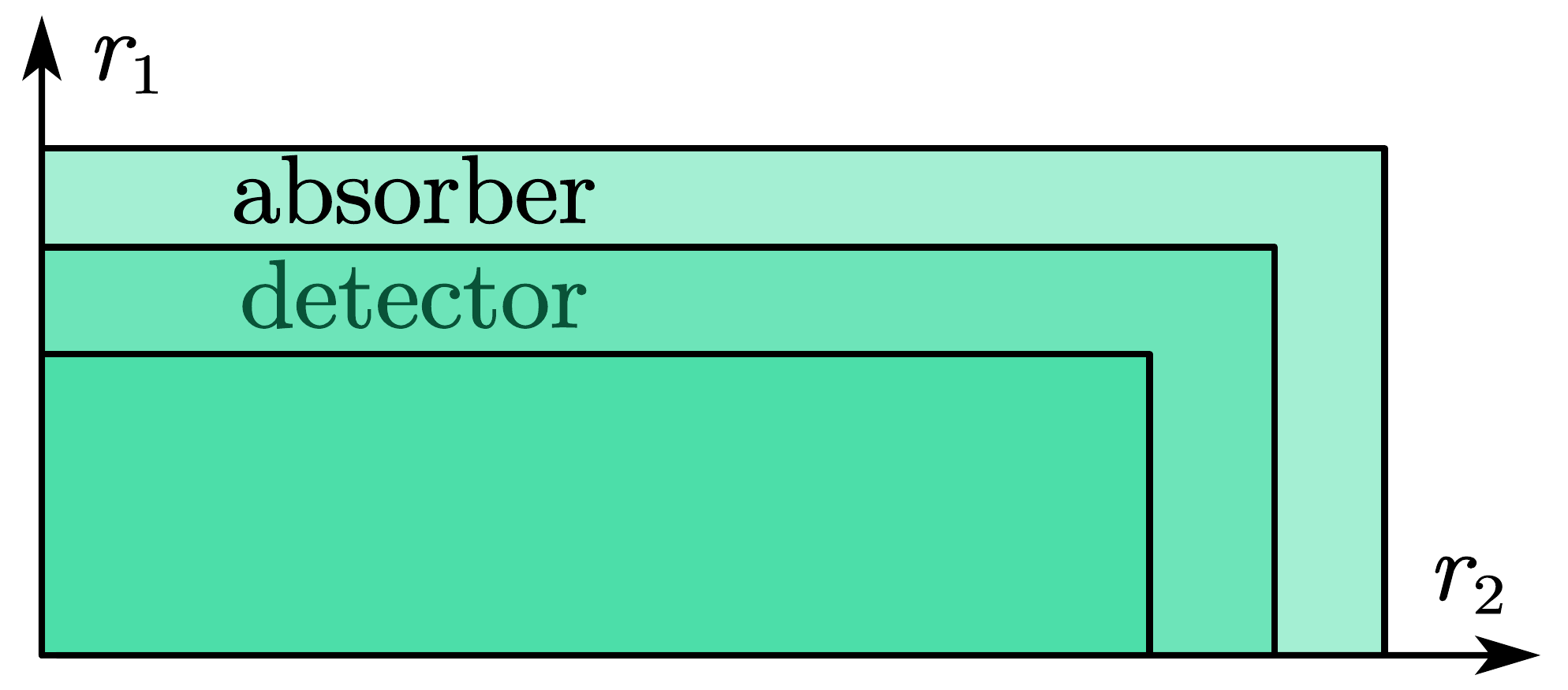}
\caption{2D grid layout for every partial waves in Eq.~\Eqref{twoelwf}.}
\label{fig:grid_layout}
\end{figure}

We propagate the radial wave function on an adaptive $(r_1, r_2$) grid, which we determine during the numerical calculation by adding a "detector region" outside the complex absorber potentials at the outer grid boundaries of $r_1$ and $r_2$~\cite{Feuerstein2003} (\Figref{grid_layout}). When the wave-function probability density in the detector region on either side reaches a threshold (typically  $5\e{-9}$ ), the length of that side is automatically extended, keeping the existing grid points and basis polynomials unchanged. For single ionization, one radial coordinate has a fixed length and no detector region is needed along its grid direction. In tests we found that this technique can accelerate the numerical throughput for single ionization 2 to 3 times. 

\vspace{1cm}
\section*{Acknowledgments}

We thank Keegan Finger, who participated in the early stages of this work within the "Research Experiences for Undergraduates (REU)" program of the NSF, as well as Van-Hung Hoang and Aihua Liu for stimulating discussions. 
This work was supported by NSF Grant ${{\text {No. 2110633}}}$ (Method development for the assessment of transient response in matter to and electronic states in arbitrary time-dependent fields)
and the Chemical Sciences, Geosciences, and Biosciences Division, Office of Basic Energy Sciences, Office of Science, US Department of Energy under Award DEFG02-86ER13491 (Attosecond electron dynamics). 

\bibliography{he_streaking.bib}

\providecommand{\noopsort}[1]{}\providecommand{\singleletter}[1]{#1}%
\begin{thebibliography}{39}%
\makeatletter
\providecommand \@ifxundefined [1]{%
 \@ifx{#1\undefined}
}%
\providecommand \@ifnum [1]{%
 \ifnum #1\expandafter \@firstoftwo
 \else \expandafter \@secondoftwo
 \fi
}%
\providecommand \@ifx [1]{%
 \ifx #1\expandafter \@firstoftwo
 \else \expandafter \@secondoftwo
 \fi
}%
\providecommand \natexlab [1]{#1}%
\providecommand \enquote  [1]{``#1''}%
\providecommand \bibnamefont  [1]{#1}%
\providecommand \bibfnamefont [1]{#1}%
\providecommand \citenamefont [1]{#1}%
\providecommand \href@noop [0]{\@secondoftwo}%
\providecommand \href [0]{\begingroup \@sanitize@url \@href}%
\providecommand \@href[1]{\@@startlink{#1}\@@href}%
\providecommand \@@href[1]{\endgroup#1\@@endlink}%
\providecommand \@sanitize@url [0]{\catcode `\\12\catcode `\$12\catcode
  `\&12\catcode `\#12\catcode `\^12\catcode `\_12\catcode `\%12\relax}%
\providecommand \@@startlink[1]{}%
\providecommand \@@endlink[0]{}%
\providecommand \url  [0]{\begingroup\@sanitize@url \@url }%
\providecommand \@url [1]{\endgroup\@href {#1}{\urlprefix }}%
\providecommand \urlprefix  [0]{URL }%
\providecommand \Eprint [0]{\href }%
\providecommand \doibase [0]{https://doi.org/}%
\providecommand \selectlanguage [0]{\@gobble}%
\providecommand \bibinfo  [0]{\@secondoftwo}%
\providecommand \bibfield  [0]{\@secondoftwo}%
\providecommand \translation [1]{[#1]}%
\providecommand \BibitemOpen [0]{}%
\providecommand \bibitemStop [0]{}%
\providecommand \bibitemNoStop [0]{.\EOS\space}%
\providecommand \EOS [0]{\spacefactor3000\relax}%
\providecommand \BibitemShut  [1]{\csname bibitem#1\endcsname}%
\let\auto@bib@innerbib\@empty
%</preamble>
\bibitem [{\citenamefont {Schultze}\ \emph {et~al.}(2010)\citenamefont
  {Schultze}, \citenamefont {Fie{\ss}}, \citenamefont {Karpowicz},
  \citenamefont {Gagnon}, \citenamefont {Korbman}, \citenamefont {Hofstetter},
  \citenamefont {Neppl}, \citenamefont {Cavalieri}, \citenamefont {Komninos},
  \citenamefont {Mercouris}, \citenamefont {Nicolaides}, \citenamefont
  {Pazourek}, \citenamefont {Nagele}, \citenamefont {Feist}, \citenamefont
  {Burgdörfer}, \citenamefont {Azzeer}, \citenamefont {Ernstorfer},
  \citenamefont {Kienberger}, \citenamefont {Kleineberg}, \citenamefont
  {Goulielmakis}, \citenamefont {Krausz},\ and\ \citenamefont
  {Yakovlev}}]{Schu10}%
  \BibitemOpen
  \bibfield  {author} {\bibinfo {author} {\bibfnamefont {M.}~\bibnamefont
  {Schultze}}, \bibinfo {author} {\bibfnamefont {M.}~\bibnamefont {Fie{\ss}}},
  \bibinfo {author} {\bibfnamefont {N.}~\bibnamefont {Karpowicz}}, \bibinfo
  {author} {\bibfnamefont {J.}~\bibnamefont {Gagnon}}, \bibinfo {author}
  {\bibfnamefont {M.}~\bibnamefont {Korbman}}, \bibinfo {author} {\bibfnamefont
  {M.}~\bibnamefont {Hofstetter}}, \bibinfo {author} {\bibfnamefont
  {S.}~\bibnamefont {Neppl}}, \bibinfo {author} {\bibfnamefont {A.~L.}\
  \bibnamefont {Cavalieri}}, \bibinfo {author} {\bibfnamefont {Y.}~\bibnamefont
  {Komninos}}, \bibinfo {author} {\bibfnamefont {T.}~\bibnamefont {Mercouris}},
  \bibinfo {author} {\bibfnamefont {C.~A.}\ \bibnamefont {Nicolaides}},
  \bibinfo {author} {\bibfnamefont {R.}~\bibnamefont {Pazourek}}, \bibinfo
  {author} {\bibfnamefont {S.}~\bibnamefont {Nagele}}, \bibinfo {author}
  {\bibfnamefont {J.}~\bibnamefont {Feist}}, \bibinfo {author} {\bibfnamefont
  {J.}~\bibnamefont {Burgdörfer}}, \bibinfo {author} {\bibfnamefont {A.~M.}\
  \bibnamefont {Azzeer}}, \bibinfo {author} {\bibfnamefont {R.}~\bibnamefont
  {Ernstorfer}}, \bibinfo {author} {\bibfnamefont {R.}~\bibnamefont
  {Kienberger}}, \bibinfo {author} {\bibfnamefont {U.}~\bibnamefont
  {Kleineberg}}, \bibinfo {author} {\bibfnamefont {E.}~\bibnamefont
  {Goulielmakis}}, \bibinfo {author} {\bibfnamefont {F.}~\bibnamefont
  {Krausz}},\ and\ \bibinfo {author} {\bibfnamefont {V.~S.}\ \bibnamefont
  {Yakovlev}},\ }\bibfield  {title} {\bibinfo {title} {Delay in
  photoemission},\ }\href {https://doi.org/10.1126/science.1189401} {\bibfield
  {journal} {\bibinfo  {journal} {Science}\ }\textbf {\bibinfo {volume}
  {328}},\ \bibinfo {pages} {1658} (\bibinfo {year} {2010})}\BibitemShut
  {NoStop}%
\bibitem [{\citenamefont {Ossiander}\ \emph {et~al.}(2017)\citenamefont
  {Ossiander}, \citenamefont {Siegrist}, \citenamefont {V.~Shirvanyan},
  \citenamefont {Sommer}, \citenamefont {Latka}, \citenamefont {Guggenmos},
  \citenamefont {Nagele}, \citenamefont {Feist}, \citenamefont {Burgdörfer},
  \citenamefont {Kienberger},\ and\ \citenamefont {Schultze}}]{Ossi17}%
  \BibitemOpen
  \bibfield  {author} {\bibinfo {author} {\bibfnamefont {M.}~\bibnamefont
  {Ossiander}}, \bibinfo {author} {\bibfnamefont {F.}~\bibnamefont {Siegrist}},
  \bibinfo {author} {\bibfnamefont {R.~P.}\ \bibnamefont {V.~Shirvanyan}},
  \bibinfo {author} {\bibfnamefont {A.}~\bibnamefont {Sommer}}, \bibinfo
  {author} {\bibfnamefont {T.}~\bibnamefont {Latka}}, \bibinfo {author}
  {\bibfnamefont {A.}~\bibnamefont {Guggenmos}}, \bibinfo {author}
  {\bibfnamefont {S.}~\bibnamefont {Nagele}}, \bibinfo {author} {\bibfnamefont
  {J.}~\bibnamefont {Feist}}, \bibinfo {author} {\bibfnamefont
  {J.}~\bibnamefont {Burgdörfer}}, \bibinfo {author} {\bibfnamefont
  {R.}~\bibnamefont {Kienberger}},\ and\ \bibinfo {author} {\bibfnamefont
  {M.}~\bibnamefont {Schultze}},\ }\bibfield  {title} {\bibinfo {title}
  {Attosecond correlation dynamics},\ }\href@noop {} {\bibfield  {journal}
  {\bibinfo  {journal} {Nat. Phys.}\ }\textbf {\bibinfo {volume} {13}},\
  \bibinfo {pages} {280} (\bibinfo {year} {2017})}\BibitemShut {NoStop}%
\bibitem [{\citenamefont {Quan}\ \emph {et~al.}(2019)\citenamefont {Quan},
  \citenamefont {Serov}, \citenamefont {Wei}, \citenamefont {Zhao},
  \citenamefont {Zhou}, \citenamefont {Wang}, \citenamefont {Lai},
  \citenamefont {Kheifets},\ and\ \citenamefont {Liu}}]{Quan19}%
  \BibitemOpen
  \bibfield  {author} {\bibinfo {author} {\bibfnamefont {W.}~\bibnamefont
  {Quan}}, \bibinfo {author} {\bibfnamefont {V.~V.}\ \bibnamefont {Serov}},
  \bibinfo {author} {\bibfnamefont {M.}~\bibnamefont {Wei}}, \bibinfo {author}
  {\bibfnamefont {M.}~\bibnamefont {Zhao}}, \bibinfo {author} {\bibfnamefont
  {Y.}~\bibnamefont {Zhou}}, \bibinfo {author} {\bibfnamefont {Y.}~\bibnamefont
  {Wang}}, \bibinfo {author} {\bibfnamefont {X.}~\bibnamefont {Lai}}, \bibinfo
  {author} {\bibfnamefont {A.~S.}\ \bibnamefont {Kheifets}},\ and\ \bibinfo
  {author} {\bibfnamefont {X.}~\bibnamefont {Liu}},\ }\bibfield  {title}
  {\bibinfo {title} {Attosecond molecular angular streaking with all-ionic
  fragments detection},\ }\href
  {https://doi.org/10.1103/PhysRevLett.123.223204} {\bibfield  {journal}
  {\bibinfo  {journal} {Phys. Rev. Lett.}\ }\textbf {\bibinfo {volume} {123}},\
  \bibinfo {pages} {223204} (\bibinfo {year} {2019})}\BibitemShut {NoStop}%
\bibitem [{\citenamefont {Cattaneo}\ \emph {et~al.}(2022)\citenamefont
  {Cattaneo}, \citenamefont {Pedrelli}, \citenamefont {Bello}, \citenamefont
  {Palacios}, \citenamefont {Keathley}, \citenamefont {Mart\'{\i}n},\ and\
  \citenamefont {Keller}}]{Catt22}%
  \BibitemOpen
  \bibfield  {author} {\bibinfo {author} {\bibfnamefont {L.}~\bibnamefont
  {Cattaneo}}, \bibinfo {author} {\bibfnamefont {L.}~\bibnamefont {Pedrelli}},
  \bibinfo {author} {\bibfnamefont {R.~Y.}\ \bibnamefont {Bello}}, \bibinfo
  {author} {\bibfnamefont {A.}~\bibnamefont {Palacios}}, \bibinfo {author}
  {\bibfnamefont {P.~D.}\ \bibnamefont {Keathley}}, \bibinfo {author}
  {\bibfnamefont {F.}~\bibnamefont {Mart\'{\i}n}},\ and\ \bibinfo {author}
  {\bibfnamefont {U.}~\bibnamefont {Keller}},\ }\bibfield  {title} {\bibinfo
  {title} {Isolating attosecond electron dynamics in molecules where nuclei
  move fast},\ }\href {https://doi.org/10.1103/PhysRevLett.128.063001}
  {\bibfield  {journal} {\bibinfo  {journal} {Phys. Rev. Lett.}\ }\textbf
  {\bibinfo {volume} {128}},\ \bibinfo {pages} {063001} (\bibinfo {year}
  {2022})}\BibitemShut {NoStop}%
\bibitem [{\citenamefont {Kowalewski}\ \emph {et~al.}(2016)\citenamefont
  {Kowalewski}, \citenamefont {Bennett}, \citenamefont {Rouxel},\ and\
  \citenamefont {Mukamel}}]{Kowa16}%
  \BibitemOpen
  \bibfield  {author} {\bibinfo {author} {\bibfnamefont {M.}~\bibnamefont
  {Kowalewski}}, \bibinfo {author} {\bibfnamefont {K.}~\bibnamefont {Bennett}},
  \bibinfo {author} {\bibfnamefont {J.~R.}\ \bibnamefont {Rouxel}},\ and\
  \bibinfo {author} {\bibfnamefont {S.}~\bibnamefont {Mukamel}},\ }\bibfield
  {title} {\bibinfo {title} {Monitoring nonadiabatic electron-nuclear dynamics
  in molecules by attosecond streaking of photoelectrons},\ }\href
  {https://doi.org/10.1103/PhysRevLett.117.043201} {\bibfield  {journal}
  {\bibinfo  {journal} {Phys. Rev. Lett.}\ }\textbf {\bibinfo {volume} {117}},\
  \bibinfo {pages} {043201} (\bibinfo {year} {2016})}\BibitemShut {NoStop}%
\bibitem [{\citenamefont {Cavalieri}\ \emph {et~al.}(2007)\citenamefont
  {Cavalieri}, \citenamefont {Müller}, \citenamefont {Uphues}, \citenamefont
  {Yakovlev}, \citenamefont {Baltuška}, \citenamefont {Horvath}, \citenamefont
  {Schmidt}, \citenamefont {Blümel}, \citenamefont {Holzwarth}, \citenamefont
  {Hendel}, \citenamefont {Drescher}, \citenamefont {Kleineberg}, \citenamefont
  {Echenique}, \citenamefont {Kienberger}, \citenamefont {Krausz},\ and\
  \citenamefont {Heinzmann}}]{Cava07}%
  \BibitemOpen
  \bibfield  {author} {\bibinfo {author} {\bibfnamefont {A.~L.}\ \bibnamefont
  {Cavalieri}}, \bibinfo {author} {\bibfnamefont {N.}~\bibnamefont {Müller}},
  \bibinfo {author} {\bibfnamefont {T.}~\bibnamefont {Uphues}}, \bibinfo
  {author} {\bibfnamefont {V.~S.}\ \bibnamefont {Yakovlev}}, \bibinfo {author}
  {\bibfnamefont {A.}~\bibnamefont {Baltuška}}, \bibinfo {author}
  {\bibfnamefont {B.}~\bibnamefont {Horvath}}, \bibinfo {author} {\bibfnamefont
  {B.}~\bibnamefont {Schmidt}}, \bibinfo {author} {\bibfnamefont
  {L.}~\bibnamefont {Blümel}}, \bibinfo {author} {\bibfnamefont
  {R.}~\bibnamefont {Holzwarth}}, \bibinfo {author} {\bibfnamefont
  {S.}~\bibnamefont {Hendel}}, \bibinfo {author} {\bibfnamefont
  {M.}~\bibnamefont {Drescher}}, \bibinfo {author} {\bibfnamefont
  {U.}~\bibnamefont {Kleineberg}}, \bibinfo {author} {\bibfnamefont {P.~M.}\
  \bibnamefont {Echenique}}, \bibinfo {author} {\bibfnamefont {R.}~\bibnamefont
  {Kienberger}}, \bibinfo {author} {\bibfnamefont {F.}~\bibnamefont {Krausz}},\
  and\ \bibinfo {author} {\bibfnamefont {U.}~\bibnamefont {Heinzmann}},\
  }\bibfield  {title} {\bibinfo {title} {Attosecond spectroscopy in condensed
  matter},\ }\href {https://doi.org/10.1038/nature06229} {\bibfield  {journal}
  {\bibinfo  {journal} {Nature}\ }\textbf {\bibinfo {volume} {449}},\ \bibinfo
  {pages} {1029} (\bibinfo {year} {2007})}\BibitemShut {NoStop}%
\bibitem [{\citenamefont {Förg}\ \emph {et~al.}(2016)\citenamefont {Förg},
  \citenamefont {Schötz}, \citenamefont {Süßmann}, \citenamefont {Förster},
  \citenamefont {Krüger}, \citenamefont {Ahn}, \citenamefont {Okell},
  \citenamefont {Wintersperger}, \citenamefont {Zherebtsov}, \citenamefont
  {Guggenmos}, \citenamefont {Pervak}, \citenamefont {Kessel}, \citenamefont
  {Trushin}, \citenamefont {Azzeer}, \citenamefont {Stockman}, \citenamefont
  {Kim}, \citenamefont {Krausz}, \citenamefont {Hommelhoff},\ and\
  \citenamefont {Kling}}]{Forg16}%
  \BibitemOpen
  \bibfield  {author} {\bibinfo {author} {\bibfnamefont {B.}~\bibnamefont
  {Förg}}, \bibinfo {author} {\bibfnamefont {J.}~\bibnamefont {Schötz}},
  \bibinfo {author} {\bibfnamefont {F.}~\bibnamefont {Süßmann}}, \bibinfo
  {author} {\bibfnamefont {M.}~\bibnamefont {Förster}}, \bibinfo {author}
  {\bibfnamefont {M.}~\bibnamefont {Krüger}}, \bibinfo {author} {\bibfnamefont
  {B.}~\bibnamefont {Ahn}}, \bibinfo {author} {\bibfnamefont {W.~A.}\
  \bibnamefont {Okell}}, \bibinfo {author} {\bibfnamefont {K.}~\bibnamefont
  {Wintersperger}}, \bibinfo {author} {\bibfnamefont {S.}~\bibnamefont
  {Zherebtsov}}, \bibinfo {author} {\bibfnamefont {A.}~\bibnamefont
  {Guggenmos}}, \bibinfo {author} {\bibfnamefont {V.}~\bibnamefont {Pervak}},
  \bibinfo {author} {\bibfnamefont {A.}~\bibnamefont {Kessel}}, \bibinfo
  {author} {\bibfnamefont {S.~A.}\ \bibnamefont {Trushin}}, \bibinfo {author}
  {\bibfnamefont {A.~M.}\ \bibnamefont {Azzeer}}, \bibinfo {author}
  {\bibfnamefont {M.~I.}\ \bibnamefont {Stockman}}, \bibinfo {author}
  {\bibfnamefont {D.}~\bibnamefont {Kim}}, \bibinfo {author} {\bibfnamefont
  {F.}~\bibnamefont {Krausz}}, \bibinfo {author} {\bibfnamefont
  {P.}~\bibnamefont {Hommelhoff}},\ and\ \bibinfo {author} {\bibfnamefont
  {M.~F.}\ \bibnamefont {Kling}},\ }\bibfield  {title} {\bibinfo {title}
  {Attosecond nanoscale near-field sampling},\ }\href
  {https://doi.org/10.1038/ncomms11717} {\bibfield  {journal} {\bibinfo
  {journal} {Nat. Commun.}\ }\textbf {\bibinfo {volume} {7}},\ \bibinfo {pages}
  {11717} (\bibinfo {year} {2016})}\BibitemShut {NoStop}%
\bibitem [{\citenamefont {Ossiander}\ \emph {et~al.}(2018)\citenamefont
  {Ossiander}, \citenamefont {Riemensberger}, \citenamefont {Neppl},
  \citenamefont {Mittermair}, \citenamefont {Schäffer}, \citenamefont
  {Duensing}, \citenamefont {Wagner}, \citenamefont {Heider}, \citenamefont
  {Wurzer}, \citenamefont {Gerl}, \citenamefont {Schnitzenbaumer},
  \citenamefont {Barth}, \citenamefont {Libisch}, \citenamefont {Lemell},
  \citenamefont {Burgdörfer}, \citenamefont {Feulner},\ and\ \citenamefont
  {Kienberger}}]{Ossi18}%
  \BibitemOpen
  \bibfield  {author} {\bibinfo {author} {\bibfnamefont {M.}~\bibnamefont
  {Ossiander}}, \bibinfo {author} {\bibfnamefont {J.}~\bibnamefont
  {Riemensberger}}, \bibinfo {author} {\bibfnamefont {S.}~\bibnamefont
  {Neppl}}, \bibinfo {author} {\bibfnamefont {M.}~\bibnamefont {Mittermair}},
  \bibinfo {author} {\bibfnamefont {M.}~\bibnamefont {Schäffer}}, \bibinfo
  {author} {\bibfnamefont {A.}~\bibnamefont {Duensing}}, \bibinfo {author}
  {\bibfnamefont {M.~S.}\ \bibnamefont {Wagner}}, \bibinfo {author}
  {\bibfnamefont {R.}~\bibnamefont {Heider}}, \bibinfo {author} {\bibfnamefont
  {M.}~\bibnamefont {Wurzer}}, \bibinfo {author} {\bibfnamefont
  {M.}~\bibnamefont {Gerl}}, \bibinfo {author} {\bibfnamefont {M.}~\bibnamefont
  {Schnitzenbaumer}}, \bibinfo {author} {\bibfnamefont {J.~V.}\ \bibnamefont
  {Barth}}, \bibinfo {author} {\bibfnamefont {F.}~\bibnamefont {Libisch}},
  \bibinfo {author} {\bibfnamefont {C.}~\bibnamefont {Lemell}}, \bibinfo
  {author} {\bibfnamefont {J.}~\bibnamefont {Burgdörfer}}, \bibinfo {author}
  {\bibfnamefont {P.}~\bibnamefont {Feulner}},\ and\ \bibinfo {author}
  {\bibfnamefont {R.}~\bibnamefont {Kienberger}},\ }\bibfield  {title}
  {\bibinfo {title} {Absolute timing of the photoelectric effect},\ }\href
  {https://doi.org/10.1038/s41586-018-0503-6} {\bibfield  {journal} {\bibinfo
  {journal} {Nature}\ }\textbf {\bibinfo {volume} {561}},\ \bibinfo {pages}
  {374} (\bibinfo {year} {2018})}\BibitemShut {NoStop}%
\bibitem [{\citenamefont {Hentschel}\ \emph {et~al.}(2001)\citenamefont
  {Hentschel}, \citenamefont {Kienberger}, \citenamefont {Spielmann},
  \citenamefont {Reider}, \citenamefont {Milosevic}, \citenamefont {Brabec},
  \citenamefont {Corkum}, \citenamefont {Heinzmann}, \citenamefont {Drescher},\
  and\ \citenamefont {Krausz}}]{Hent01}%
  \BibitemOpen
  \bibfield  {author} {\bibinfo {author} {\bibfnamefont {M.}~\bibnamefont
  {Hentschel}}, \bibinfo {author} {\bibfnamefont {R.}~\bibnamefont
  {Kienberger}}, \bibinfo {author} {\bibfnamefont {C.}~\bibnamefont
  {Spielmann}}, \bibinfo {author} {\bibfnamefont {G.~A.}\ \bibnamefont
  {Reider}}, \bibinfo {author} {\bibfnamefont {N.}~\bibnamefont {Milosevic}},
  \bibinfo {author} {\bibfnamefont {T.}~\bibnamefont {Brabec}}, \bibinfo
  {author} {\bibfnamefont {P.}~\bibnamefont {Corkum}}, \bibinfo {author}
  {\bibfnamefont {U.}~\bibnamefont {Heinzmann}}, \bibinfo {author}
  {\bibfnamefont {M.}~\bibnamefont {Drescher}},\ and\ \bibinfo {author}
  {\bibfnamefont {F.}~\bibnamefont {Krausz}},\ }\bibfield  {title} {\bibinfo
  {title} {Attosecond metrology},\ }\href {https://doi.org/10.1038/35107000}
  {\bibfield  {journal} {\bibinfo  {journal} {Nature}\ }\textbf {\bibinfo
  {volume} {414}},\ \bibinfo {pages} {509} (\bibinfo {year}
  {2001})}\BibitemShut {NoStop}%
\bibitem [{\citenamefont {Goulielmakis}\ \emph {et~al.}(2004)\citenamefont
  {Goulielmakis}, \citenamefont {Uiberacker}, \citenamefont {Kienberger},
  \citenamefont {Baltuska}, \citenamefont {Yakovlev}, \citenamefont {Scrinzi},
  \citenamefont {Westerwalbesloh}, \citenamefont {Kleineberg}, \citenamefont
  {Heinzmann}, \citenamefont {Drescher},\ and\ \citenamefont
  {Krausz}}]{Goul04}%
  \BibitemOpen
  \bibfield  {author} {\bibinfo {author} {\bibfnamefont {E.}~\bibnamefont
  {Goulielmakis}}, \bibinfo {author} {\bibfnamefont {M.}~\bibnamefont
  {Uiberacker}}, \bibinfo {author} {\bibfnamefont {R.}~\bibnamefont
  {Kienberger}}, \bibinfo {author} {\bibfnamefont {A.}~\bibnamefont
  {Baltuska}}, \bibinfo {author} {\bibfnamefont {V.}~\bibnamefont {Yakovlev}},
  \bibinfo {author} {\bibfnamefont {A.}~\bibnamefont {Scrinzi}}, \bibinfo
  {author} {\bibfnamefont {T.}~\bibnamefont {Westerwalbesloh}}, \bibinfo
  {author} {\bibfnamefont {U.}~\bibnamefont {Kleineberg}}, \bibinfo {author}
  {\bibfnamefont {U.}~\bibnamefont {Heinzmann}}, \bibinfo {author}
  {\bibfnamefont {M.}~\bibnamefont {Drescher}},\ and\ \bibinfo {author}
  {\bibfnamefont {F.}~\bibnamefont {Krausz}},\ }\bibfield  {title} {\bibinfo
  {title} {Direct measurement of light waves},\ }\href
  {https://doi.org/10.1126/science.1100866} {\bibfield  {journal} {\bibinfo
  {journal} {Science}\ }\textbf {\bibinfo {volume} {305}},\ \bibinfo {pages}
  {1267} (\bibinfo {year} {2004})}\BibitemShut {NoStop}%
\bibitem [{\citenamefont {Thumm}\ \emph {et~al.}(2015)\citenamefont {Thumm},
  \citenamefont {Liao}, \citenamefont {Bothschafter}, \citenamefont
  {S\"u\ss{}mann}, \citenamefont {Kling},\ and\ \citenamefont
  {Kienberger}}]{ThummChap15}%
  \BibitemOpen
  \bibfield  {author} {\bibinfo {author} {\bibfnamefont {U.}~\bibnamefont
  {Thumm}}, \bibinfo {author} {\bibfnamefont {Q.}~\bibnamefont {Liao}},
  \bibinfo {author} {\bibfnamefont {E.~M.}\ \bibnamefont {Bothschafter}},
  \bibinfo {author} {\bibfnamefont {F.}~\bibnamefont {S\"u\ss{}mann}}, \bibinfo
  {author} {\bibfnamefont {M.~F.}\ \bibnamefont {Kling}},\ and\ \bibinfo
  {author} {\bibfnamefont {R.}~\bibnamefont {Kienberger}},\ }\bibfield  {title}
  {\bibinfo {title} {Fundamentals of photonics and physics},\ }in\ \href@noop
  {} {\emph {\bibinfo {booktitle} {The Oxford Handbook of Innovation}}},\
  \bibinfo {editor} {edited by\ \bibinfo {editor} {\bibfnamefont
  {D.}~\bibnamefont {Andrew}}}\ (\bibinfo  {publisher} {Wiley},\ \bibinfo
  {address} {New York},\ \bibinfo {year} {2015})\ Chap.~\bibinfo {chapter}
  {13}\BibitemShut {NoStop}%
\bibitem [{\citenamefont {Kheifets}\ and\ \citenamefont
  {Ivanov}(2010)}]{Khei10}%
  \BibitemOpen
  \bibfield  {author} {\bibinfo {author} {\bibfnamefont {A.~S.}\ \bibnamefont
  {Kheifets}}\ and\ \bibinfo {author} {\bibfnamefont {I.~A.}\ \bibnamefont
  {Ivanov}},\ }\bibfield  {title} {\bibinfo {title} {Delay in atomic
  photoionization},\ }\href@noop {} {\bibfield  {journal} {\bibinfo  {journal}
  {Phys. Rev. Lett.}\ }\textbf {\bibinfo {volume} {105}},\ \bibinfo {pages}
  {233002} (\bibinfo {year} {2010})}\BibitemShut {NoStop}%
\bibitem [{\citenamefont {Moore}\ \emph {et~al.}(2011)\citenamefont {Moore},
  \citenamefont {Lysaght}, \citenamefont {Parker}, \citenamefont {van~der
  Hart},\ and\ \citenamefont {Taylor}}]{Moor11}%
  \BibitemOpen
  \bibfield  {author} {\bibinfo {author} {\bibfnamefont {L.~R.}\ \bibnamefont
  {Moore}}, \bibinfo {author} {\bibfnamefont {M.~A.}\ \bibnamefont {Lysaght}},
  \bibinfo {author} {\bibfnamefont {J.~S.}\ \bibnamefont {Parker}}, \bibinfo
  {author} {\bibfnamefont {H.~W.}\ \bibnamefont {van~der Hart}},\ and\ \bibinfo
  {author} {\bibfnamefont {K.~T.}\ \bibnamefont {Taylor}},\ }\bibfield  {title}
  {\bibinfo {title} {Time delay between photoemission from the $2p$ and $2s$
  subshells of neon},\ }\href@noop {} {\bibfield  {journal} {\bibinfo
  {journal} {Phys. Rev. A}\ }\textbf {\bibinfo {volume} {84}},\ \bibinfo
  {pages} {061404} (\bibinfo {year} {2011})}\BibitemShut {NoStop}%
\bibitem [{\citenamefont {Nagele}\ \emph {et~al.}(2012)\citenamefont {Nagele},
  \citenamefont {Pazourek}, \citenamefont {Feist},\ and\ \citenamefont
  {Burgdörfer}}]{Nage12}%
  \BibitemOpen
  \bibfield  {author} {\bibinfo {author} {\bibfnamefont {S.}~\bibnamefont
  {Nagele}}, \bibinfo {author} {\bibfnamefont {R.}~\bibnamefont {Pazourek}},
  \bibinfo {author} {\bibfnamefont {J.}~\bibnamefont {Feist}},\ and\ \bibinfo
  {author} {\bibfnamefont {J.}~\bibnamefont {Burgdörfer}},\ }\bibfield
  {title} {\bibinfo {title} {Time shifts in photoemission from a fully
  correlated two-electron model system},\ }\href
  {https://doi.org/10.1103/PhysRevA.85.033401} {\bibfield  {journal} {\bibinfo
  {journal} {Phys. Rev. A}\ }\textbf {\bibinfo {volume} {85}},\ \bibinfo
  {pages} {033401} (\bibinfo {year} {2012})}\BibitemShut {NoStop}%
\bibitem [{\citenamefont {Feist}\ \emph {et~al.}(2014)\citenamefont {Feist},
  \citenamefont {Zatsarinny}, \citenamefont {Nagele}, \citenamefont {Pazourek},
  \citenamefont {Burgd\"orfer}, \citenamefont {Guan}, \citenamefont
  {Bartschat},\ and\ \citenamefont {Schneider}}]{Feis14}%
  \BibitemOpen
  \bibfield  {author} {\bibinfo {author} {\bibfnamefont {J.}~\bibnamefont
  {Feist}}, \bibinfo {author} {\bibfnamefont {O.}~\bibnamefont {Zatsarinny}},
  \bibinfo {author} {\bibfnamefont {S.}~\bibnamefont {Nagele}}, \bibinfo
  {author} {\bibfnamefont {R.}~\bibnamefont {Pazourek}}, \bibinfo {author}
  {\bibfnamefont {J.}~\bibnamefont {Burgd\"orfer}}, \bibinfo {author}
  {\bibfnamefont {X.}~\bibnamefont {Guan}}, \bibinfo {author} {\bibfnamefont
  {K.}~\bibnamefont {Bartschat}},\ and\ \bibinfo {author} {\bibfnamefont
  {B.~I.}\ \bibnamefont {Schneider}},\ }\bibfield  {title} {\bibinfo {title}
  {Time delays for attosecond streaking in photoionization of neon},\ }\href
  {https://doi.org/10.1103/PhysRevA.89.033417} {\bibfield  {journal} {\bibinfo
  {journal} {Phys. Rev. A}\ }\textbf {\bibinfo {volume} {89}},\ \bibinfo
  {pages} {033417} (\bibinfo {year} {2014})}\BibitemShut {NoStop}%
\bibitem [{\citenamefont {Dahlström}\ \emph {et~al.}(2012)\citenamefont
  {Dahlström}, \citenamefont {Carette},\ and\ \citenamefont
  {Lindroth}}]{Dahl12b}%
  \BibitemOpen
  \bibfield  {author} {\bibinfo {author} {\bibfnamefont {J.~M.}\ \bibnamefont
  {Dahlström}}, \bibinfo {author} {\bibfnamefont {T.}~\bibnamefont
  {Carette}},\ and\ \bibinfo {author} {\bibfnamefont {E.}~\bibnamefont
  {Lindroth}},\ }\bibfield  {title} {\bibinfo {title} {Diagrammatic approach to
  attosecond delays in photoionization},\ }\href@noop {} {\bibfield  {journal}
  {\bibinfo  {journal} {Phys. Rev. A}\ }\textbf {\bibinfo {volume} {86}},\
  \bibinfo {pages} {061402} (\bibinfo {year} {2012})}\BibitemShut {NoStop}%
\bibitem [{\citenamefont {Kheifets}(2023)}]{Khei23}%
  \BibitemOpen
  \bibfield  {author} {\bibinfo {author} {\bibfnamefont {A.~S.}\ \bibnamefont
  {Kheifets}},\ }\bibfield  {title} {\bibinfo {title} {Wigner time delay in
  atomic photoionization},\ }\href {https://doi.org/10.1088/1361-6455/acb188}
  {\bibfield  {journal} {\bibinfo  {journal} {J. Phys. B}\ }\textbf {\bibinfo
  {volume} {56}},\ \bibinfo {pages} {022001} (\bibinfo {year}
  {2023})}\BibitemShut {NoStop}%
\bibitem [{\citenamefont {Isinger}\ \emph {et~al.}(2017)\citenamefont
  {Isinger}, \citenamefont {Squibb}, \citenamefont {Busto}, \citenamefont
  {Zhong}, \citenamefont {Harth}, \citenamefont {Kroon}, \citenamefont {Nandi},
  \citenamefont {Arnold}, \citenamefont {Miranda}, \citenamefont {Dahlström},
  \citenamefont {Lindroth}, \citenamefont {Feifel}, \citenamefont
  {Gisselbrecht},\ and\ \citenamefont {L’Huillier}}]{Isin17}%
  \BibitemOpen
  \bibfield  {author} {\bibinfo {author} {\bibfnamefont {M.}~\bibnamefont
  {Isinger}}, \bibinfo {author} {\bibfnamefont {R.~J.}\ \bibnamefont {Squibb}},
  \bibinfo {author} {\bibfnamefont {D.}~\bibnamefont {Busto}}, \bibinfo
  {author} {\bibfnamefont {S.}~\bibnamefont {Zhong}}, \bibinfo {author}
  {\bibfnamefont {A.}~\bibnamefont {Harth}}, \bibinfo {author} {\bibfnamefont
  {D.}~\bibnamefont {Kroon}}, \bibinfo {author} {\bibfnamefont
  {S.}~\bibnamefont {Nandi}}, \bibinfo {author} {\bibfnamefont {C.~L.}\
  \bibnamefont {Arnold}}, \bibinfo {author} {\bibfnamefont {M.}~\bibnamefont
  {Miranda}}, \bibinfo {author} {\bibfnamefont {J.~M.}\ \bibnamefont
  {Dahlström}}, \bibinfo {author} {\bibfnamefont {E.}~\bibnamefont
  {Lindroth}}, \bibinfo {author} {\bibfnamefont {R.}~\bibnamefont {Feifel}},
  \bibinfo {author} {\bibfnamefont {M.}~\bibnamefont {Gisselbrecht}},\ and\
  \bibinfo {author} {\bibfnamefont {A.}~\bibnamefont {L’Huillier}},\
  }\bibfield  {title} {\bibinfo {title} {Photoionization in the time and
  frequency domain},\ }\href@noop {} {\bibfield  {journal} {\bibinfo  {journal}
  {Science}\ }\textbf {\bibinfo {volume} {358}},\ \bibinfo {pages} {893}
  (\bibinfo {year} {2017})}\BibitemShut {NoStop}%
\bibitem [{\citenamefont {Pazourek}\ \emph {et~al.}(2012)\citenamefont
  {Pazourek}, \citenamefont {Feist}, \citenamefont {Nagele},\ and\
  \citenamefont {Burgdörfer}}]{Pazo12}%
  \BibitemOpen
  \bibfield  {author} {\bibinfo {author} {\bibfnamefont {R.}~\bibnamefont
  {Pazourek}}, \bibinfo {author} {\bibfnamefont {J.}~\bibnamefont {Feist}},
  \bibinfo {author} {\bibfnamefont {S.}~\bibnamefont {Nagele}},\ and\ \bibinfo
  {author} {\bibfnamefont {J.}~\bibnamefont {Burgdörfer}},\ }\bibfield
  {title} {\bibinfo {title} {Attosecond streaking of correlated two-electron
  transitions in helium},\ }\href
  {https://doi.org/10.1103/PhysRevLett.108.163001} {\bibfield  {journal}
  {\bibinfo  {journal} {Phys. Rev. Lett.}\ }\textbf {\bibinfo {volume} {108}},\
  \bibinfo {pages} {163001} (\bibinfo {year} {2012})}\BibitemShut {NoStop}%
\bibitem [{\citenamefont {McCurdy}\ \emph {et~al.}(2004)\citenamefont
  {McCurdy}, \citenamefont {Baertschy},\ and\ \citenamefont
  {Rescigno}}]{McCu04}%
  \BibitemOpen
  \bibfield  {author} {\bibinfo {author} {\bibfnamefont {C.~W.}\ \bibnamefont
  {McCurdy}}, \bibinfo {author} {\bibfnamefont {M.}~\bibnamefont {Baertschy}},\
  and\ \bibinfo {author} {\bibfnamefont {T.~N.}\ \bibnamefont {Rescigno}},\
  }\bibfield  {title} {\bibinfo {title} {Solving the three-body {Coulomb}
  breakup problem using exterior complex scaling},\ }\href
  {https://doi.org/10.1088/0953-4075/37/17/r01} {\bibfield  {journal} {\bibinfo
   {journal} {J. Phys. B}\ }\textbf {\bibinfo {volume} {37}},\ \bibinfo {pages}
  {R137} (\bibinfo {year} {2004})}\BibitemShut {NoStop}%
\bibitem [{\citenamefont {Zhang}\ and\ \citenamefont {Thumm}(2010)}]{Zhan10}%
  \BibitemOpen
  \bibfield  {author} {\bibinfo {author} {\bibfnamefont {C.-H.}\ \bibnamefont
  {Zhang}}\ and\ \bibinfo {author} {\bibfnamefont {U.}~\bibnamefont {Thumm}},\
  }\bibfield  {title} {\bibinfo {title} {Electron-ion interaction effects in
  attosecond time-resolved photoelectron spectra},\ }\href
  {https://doi.org/10.1103/PhysRevA.82.043405} {\bibfield  {journal} {\bibinfo
  {journal} {Phys. Rev. A}\ }\textbf {\bibinfo {volume} {82}},\ \bibinfo
  {pages} {043405} (\bibinfo {year} {2010})}\BibitemShut {NoStop}%
\bibitem [{\citenamefont {Heuser}\ \emph {et~al.}(2016)\citenamefont {Heuser},
  \citenamefont {Álvaro Jiménez~Galán}, \citenamefont {Cirelli},
  \citenamefont {Marante}, \citenamefont {Sabbar}, \citenamefont {Boge},
  \citenamefont {Lucchini}, \citenamefont {Gallmann}, \citenamefont {Ivanov},
  \citenamefont {Kheifets}, \citenamefont {Dahlström}, \citenamefont
  {Lindroth}, \citenamefont {Argenti}, \citenamefont {Martín},\ and\
  \citenamefont {Keller}}]{Heus16}%
  \BibitemOpen
  \bibfield  {author} {\bibinfo {author} {\bibfnamefont {S.}~\bibnamefont
  {Heuser}}, \bibinfo {author} {\bibnamefont {Álvaro Jiménez~Galán}},
  \bibinfo {author} {\bibfnamefont {C.}~\bibnamefont {Cirelli}}, \bibinfo
  {author} {\bibfnamefont {C.}~\bibnamefont {Marante}}, \bibinfo {author}
  {\bibfnamefont {M.}~\bibnamefont {Sabbar}}, \bibinfo {author} {\bibfnamefont
  {R.}~\bibnamefont {Boge}}, \bibinfo {author} {\bibfnamefont {M.}~\bibnamefont
  {Lucchini}}, \bibinfo {author} {\bibfnamefont {L.}~\bibnamefont {Gallmann}},
  \bibinfo {author} {\bibfnamefont {I.}~\bibnamefont {Ivanov}}, \bibinfo
  {author} {\bibfnamefont {A.~S.}\ \bibnamefont {Kheifets}}, \bibinfo {author}
  {\bibfnamefont {J.~M.}\ \bibnamefont {Dahlström}}, \bibinfo {author}
  {\bibfnamefont {E.}~\bibnamefont {Lindroth}}, \bibinfo {author}
  {\bibfnamefont {L.}~\bibnamefont {Argenti}}, \bibinfo {author} {\bibfnamefont
  {F.}~\bibnamefont {Martín}},\ and\ \bibinfo {author} {\bibfnamefont
  {U.}~\bibnamefont {Keller}},\ }\bibfield  {title} {\bibinfo {title} {Angular
  dependence of photoemission time delay in helium},\ }\href@noop {} {\bibfield
   {journal} {\bibinfo  {journal} {Phys. Rev. A}\ }\textbf {\bibinfo {volume}
  {94}},\ \bibinfo {pages} {063409} (\bibinfo {year} {2016})}\BibitemShut
  {NoStop}%
\bibitem [{\citenamefont {Fuchs}\ \emph {et~al.}(2020)\citenamefont {Fuchs},
  \citenamefont {Douguet}, \citenamefont {Donsa}, \citenamefont {Martin},
  \citenamefont {Burgdörfer}, \citenamefont {Argenti}, \citenamefont
  {Cattaneo},\ and\ \citenamefont {Keller}}]{Fuch20}%
  \BibitemOpen
  \bibfield  {author} {\bibinfo {author} {\bibfnamefont {J.}~\bibnamefont
  {Fuchs}}, \bibinfo {author} {\bibfnamefont {N.}~\bibnamefont {Douguet}},
  \bibinfo {author} {\bibfnamefont {S.}~\bibnamefont {Donsa}}, \bibinfo
  {author} {\bibfnamefont {F.}~\bibnamefont {Martin}}, \bibinfo {author}
  {\bibfnamefont {J.}~\bibnamefont {Burgdörfer}}, \bibinfo {author}
  {\bibfnamefont {L.}~\bibnamefont {Argenti}}, \bibinfo {author} {\bibfnamefont
  {L.}~\bibnamefont {Cattaneo}},\ and\ \bibinfo {author} {\bibfnamefont
  {U.}~\bibnamefont {Keller}},\ }\bibfield  {title} {\bibinfo {title} {Time
  delays from one-photon transitions in the continuum},\ }\href
  {https://doi.org/10.1364/OPTICA.378639} {\bibfield  {journal} {\bibinfo
  {journal} {Optica}\ }\textbf {\bibinfo {volume} {7}},\ \bibinfo {pages} {154}
  (\bibinfo {year} {2020})}\BibitemShut {NoStop}%
\bibitem [{\citenamefont {Liu}\ and\ \citenamefont {Thumm}(2015)}]{Liu15}%
  \BibitemOpen
  \bibfield  {author} {\bibinfo {author} {\bibfnamefont {A.}~\bibnamefont
  {Liu}}\ and\ \bibinfo {author} {\bibfnamefont {U.}~\bibnamefont {Thumm}},\
  }\bibfield  {title} {\bibinfo {title} {Criterion for distinguishing
  sequential from nonsequential contributions to the double ionization of
  helium in ultrashort extreme-ultraviolet pulses},\ }\href
  {https://doi.org/10.1103/PhysRevLett.115.183002} {\bibfield  {journal}
  {\bibinfo  {journal} {Phys. Rev. Lett.}\ }\textbf {\bibinfo {volume} {115}},\
  \bibinfo {pages} {183002} (\bibinfo {year} {2015})}\BibitemShut {NoStop}%
\bibitem [{\citenamefont {Bransden}\ and\ \citenamefont
  {Joachain}(2003)}]{Bransden}%
  \BibitemOpen
  \bibfield  {author} {\bibinfo {author} {\bibfnamefont {B.}~\bibnamefont
  {Bransden}}\ and\ \bibinfo {author} {\bibfnamefont {C.}~\bibnamefont
  {Joachain}},\ }\href@noop {} {\emph {\bibinfo {title} {Physics of atoms and
  molecules}}},\ \bibinfo {edition} {2nd}\ ed.\ (\bibinfo  {publisher}
  {Addison-Wesley},\ \bibinfo {year} {2003})\BibitemShut {NoStop}%
\bibitem [{\citenamefont {Wigner}(1955)}]{Wign55}%
  \BibitemOpen
  \bibfield  {author} {\bibinfo {author} {\bibfnamefont {E.~P.}\ \bibnamefont
  {Wigner}},\ }\bibfield  {title} {\bibinfo {title} {Lower limit for the energy
  derivative of the scattering phase shift},\ }\href@noop {} {\bibfield
  {journal} {\bibinfo  {journal} {Phys. Rev.}\ }\textbf {\bibinfo {volume}
  {98}},\ \bibinfo {pages} {145} (\bibinfo {year} {1955})}\BibitemShut
  {NoStop}%
\bibitem [{\citenamefont {de~Carvalho}\ and\ \citenamefont
  {Nussenzweig}(2002)}]{Carv02}%
  \BibitemOpen
  \bibfield  {author} {\bibinfo {author} {\bibfnamefont {C.}~\bibnamefont
  {de~Carvalho}}\ and\ \bibinfo {author} {\bibfnamefont {H.~M.}\ \bibnamefont
  {Nussenzweig}},\ }\bibfield  {title} {\bibinfo {title} {Time delay},\
  }\href@noop {} {\bibfield  {journal} {\bibinfo  {journal} {Phys. Rep.}\
  }\textbf {\bibinfo {volume} {364}},\ \bibinfo {pages} {83} (\bibinfo {year}
  {2002})}\BibitemShut {NoStop}%
\bibitem [{\citenamefont {Liao}\ and\ \citenamefont {Thumm}(2015)}]{Liao15}%
  \BibitemOpen
  \bibfield  {author} {\bibinfo {author} {\bibfnamefont {Q.}~\bibnamefont
  {Liao}}\ and\ \bibinfo {author} {\bibfnamefont {U.}~\bibnamefont {Thumm}},\
  }\bibfield  {title} {\bibinfo {title} {{Attosecond time-resolved streaked
  photoemission from Mg-covered W(110) surfaces}},\ }\href@noop {} {\bibfield
  {journal} {\bibinfo  {journal} {Phys. Rev. A}\ }\textbf {\bibinfo {volume}
  {92}},\ \bibinfo {pages} {031401} (\bibinfo {year} {2015})}\BibitemShut
  {NoStop}%
\bibitem [{\citenamefont {Pazourek}\ \emph {et~al.}(2015)\citenamefont
  {Pazourek}, \citenamefont {Nagele},\ and\ \citenamefont
  {Burgdörfer}}]{Pazo15}%
  \BibitemOpen
  \bibfield  {author} {\bibinfo {author} {\bibfnamefont {R.}~\bibnamefont
  {Pazourek}}, \bibinfo {author} {\bibfnamefont {S.}~\bibnamefont {Nagele}},\
  and\ \bibinfo {author} {\bibfnamefont {J.}~\bibnamefont {Burgdörfer}},\
  }\bibfield  {title} {\bibinfo {title} {Attosecond chronoscopy of
  photoemission},\ }\href {https://doi.org/10.1103/RevModPhys.87.765}
  {\bibfield  {journal} {\bibinfo  {journal} {Rev. Mod. Phys.}\ }\textbf
  {\bibinfo {volume} {87}},\ \bibinfo {pages} {765} (\bibinfo {year}
  {2015})}\BibitemShut {NoStop}%
\bibitem [{\citenamefont {Nagele}\ \emph {et~al.}(2011)\citenamefont {Nagele},
  \citenamefont {Pazourek}, \citenamefont {Feist}, \citenamefont
  {Doblhoff-Dier}, \citenamefont {Lemell}, \citenamefont {T{\H{o}}k{\'{e}}si},\
  and\ \citenamefont {Burgdörfer}}]{Nage11}%
  \BibitemOpen
  \bibfield  {author} {\bibinfo {author} {\bibfnamefont {S.}~\bibnamefont
  {Nagele}}, \bibinfo {author} {\bibfnamefont {R.}~\bibnamefont {Pazourek}},
  \bibinfo {author} {\bibfnamefont {J.}~\bibnamefont {Feist}}, \bibinfo
  {author} {\bibfnamefont {K.}~\bibnamefont {Doblhoff-Dier}}, \bibinfo {author}
  {\bibfnamefont {C.}~\bibnamefont {Lemell}}, \bibinfo {author} {\bibfnamefont
  {K.}~\bibnamefont {T{\H{o}}k{\'{e}}si}},\ and\ \bibinfo {author}
  {\bibfnamefont {J.}~\bibnamefont {Burgdörfer}},\ }\bibfield  {title}
  {\bibinfo {title} {Time-resolved photoemission by attosecond streaking:
  extraction of time information},\ }\href
  {https://doi.org/10.1088/0953-4075/44/8/081001} {\bibfield  {journal}
  {\bibinfo  {journal} {J. Phys. B}\ }\textbf {\bibinfo {volume} {44}},\
  \bibinfo {pages} {081001} (\bibinfo {year} {2011})}\BibitemShut {NoStop}%
\bibitem [{\citenamefont {Baggesen}\ and\ \citenamefont
  {Madsen}(2010)}]{Bagg10}%
  \BibitemOpen
  \bibfield  {author} {\bibinfo {author} {\bibfnamefont {J.~C.}\ \bibnamefont
  {Baggesen}}\ and\ \bibinfo {author} {\bibfnamefont {L.~B.}\ \bibnamefont
  {Madsen}},\ }\bibfield  {title} {\bibinfo {title} {Polarization effects in
  attosecond photoelectron spectroscopy},\ }\href
  {https://doi.org/10.1103/PhysRevLett.104.043602} {\bibfield  {journal}
  {\bibinfo  {journal} {Phys. Rev. Lett.}\ }\textbf {\bibinfo {volume} {104}},\
  \bibinfo {pages} {043602} (\bibinfo {year} {2010})}\BibitemShut {NoStop}%
\bibitem [{\citenamefont {Liu}\ and\ \citenamefont {Thumm}(2014)}]{Liu14}%
  \BibitemOpen
  \bibfield  {author} {\bibinfo {author} {\bibfnamefont {A.}~\bibnamefont
  {Liu}}\ and\ \bibinfo {author} {\bibfnamefont {U.}~\bibnamefont {Thumm}},\
  }\bibfield  {title} {\bibinfo {title} {Laser-assisted xuv few-photon double
  ionization of helium: Joint angular distributions},\ }\href
  {https://doi.org/10.1103/PhysRevA.89.063423} {\bibfield  {journal} {\bibinfo
  {journal} {Phys. Rev. A}\ }\textbf {\bibinfo {volume} {89}},\ \bibinfo
  {pages} {063423} (\bibinfo {year} {2014})}\BibitemShut {NoStop}%
\bibitem [{\citenamefont {Bandrauk}\ and\ \citenamefont {Shen}(1994)}]{Band94}%
  \BibitemOpen
  \bibfield  {author} {\bibinfo {author} {\bibfnamefont {A.~D.}\ \bibnamefont
  {Bandrauk}}\ and\ \bibinfo {author} {\bibfnamefont {H.}~\bibnamefont
  {Shen}},\ }\bibfield  {title} {\bibinfo {title} {High-order split-step
  exponential methods for solving coupled nonlinear {Schrödinger} equations},\
  }\href@noop {} {\bibfield  {journal} {\bibinfo  {journal} {J. Phys. A}\
  }\textbf {\bibinfo {volume} {27}},\ \bibinfo {pages} {7147} (\bibinfo {year}
  {1994})}\BibitemShut {NoStop}%
\bibitem [{\citenamefont {Press}\ \emph {et~al.}(2007)\citenamefont {Press},
  \citenamefont {Teukolsky}, \citenamefont {Vetterling},\ and\ \citenamefont
  {Flannery}}]{NR3}%
  \BibitemOpen
  \bibfield  {author} {\bibinfo {author} {\bibfnamefont {W.~H.}\ \bibnamefont
  {Press}}, \bibinfo {author} {\bibfnamefont {S.~A.}\ \bibnamefont
  {Teukolsky}}, \bibinfo {author} {\bibfnamefont {W.~T.}\ \bibnamefont
  {Vetterling}},\ and\ \bibinfo {author} {\bibfnamefont {B.~P.}\ \bibnamefont
  {Flannery}},\ }\href@noop {} {\emph {\bibinfo {title} {Numerical recipes: The
  art of scientific computing}}},\ \bibinfo {edition} {3rd}\ ed.\ (\bibinfo
  {publisher} {Cambridge University Press},\ \bibinfo {year}
  {2007})\BibitemShut {NoStop}%
\bibitem [{\citenamefont {Cormier}\ and\ \citenamefont
  {Lambropoulos}(1996)}]{Corm96}%
  \BibitemOpen
  \bibfield  {author} {\bibinfo {author} {\bibfnamefont {E.}~\bibnamefont
  {Cormier}}\ and\ \bibinfo {author} {\bibfnamefont {P.}~\bibnamefont
  {Lambropoulos}},\ }\bibfield  {title} {\bibinfo {title} {Optimal gauge and
  gauge invariance in nonperturbative time-dependent calculation of
  abovethreshold ionization},\ }\href@noop {} {\bibfield  {journal} {\bibinfo
  {journal} {J. Phys. B}\ }\textbf {\bibinfo {volume} {29}},\ \bibinfo {pages}
  {1667} (\bibinfo {year} {1996})}\BibitemShut {NoStop}%
\bibitem [{\citenamefont {Baye}(2012)}]{Baye12}%
  \BibitemOpen
  \bibfield  {author} {\bibinfo {author} {\bibfnamefont {D.}~\bibnamefont
  {Baye}},\ }\bibfield  {title} {\bibinfo {title} {{Exact nonrelativistic
  polarizabilities of the hydrogen atom with the Lagrange-mesh method}},\
  }\href {https://doi.org/10.1103/PhysRevA.86.062514} {\bibfield  {journal}
  {\bibinfo  {journal} {Phys. Rev. A}\ }\textbf {\bibinfo {volume} {86}},\
  \bibinfo {pages} {062514} (\bibinfo {year} {2012})}\BibitemShut {NoStop}%
\bibitem [{\citenamefont {Drake}(2006)}]{Drak06}%
  \BibitemOpen
  \bibfield  {author} {\bibinfo {author} {\bibfnamefont {G.}~\bibnamefont
  {Drake}},\ }\href@noop {} {\emph {\bibinfo {title} {Springer handbook of
  atomic, molecular, and optical physics}}}\ (\bibinfo  {publisher}
  {Springer},\ \bibinfo {year} {2006})\BibitemShut {NoStop}%
\bibitem [{\citenamefont {Bauer}\ and\ \citenamefont {Koval}(2006)}]{Baue06}%
  \BibitemOpen
  \bibfield  {author} {\bibinfo {author} {\bibfnamefont {D.}~\bibnamefont
  {Bauer}}\ and\ \bibinfo {author} {\bibfnamefont {P.}~\bibnamefont {Koval}},\
  }\bibfield  {title} {\bibinfo {title} {Qprop: A {Schrödinger}-solver for
  intense laser–atom interaction},\ }\href
  {https://doi.org/10.1016/j.cpc.2005.11.001} {\bibfield  {journal} {\bibinfo
  {journal} {Comput. Phys. Commun.}\ }\textbf {\bibinfo {volume} {174}},\
  \bibinfo {pages} {396} (\bibinfo {year} {2006})}\BibitemShut {NoStop}%
\bibitem [{\citenamefont {Feuerstein}\ and\ \citenamefont
  {Thumm}(2003)}]{Feuerstein2003}%
  \BibitemOpen
  \bibfield  {author} {\bibinfo {author} {\bibfnamefont {B.}~\bibnamefont
  {Feuerstein}}\ and\ \bibinfo {author} {\bibfnamefont {U.}~\bibnamefont
  {Thumm}},\ }\bibfield  {title} {\bibinfo {title} {On the computation of
  momentum distributions within wavepacket propagation calculations},\ }\href
  {https://doi.org/10.1103/PhysRevLett.104.043602} {\bibfield  {journal}
  {\bibinfo  {journal} {J. Phys. B}\ }\textbf {\bibinfo {volume} {36}},\
  \bibinfo {pages} {707} (\bibinfo {year} {2003})}\BibitemShut {NoStop}%
\end{thebibliography}%

\end{document}